\documentclass[twocolumn,trackchanges]{aastex701}

\usepackage{natbib}
\setcitestyle{aysep={}}
\usepackage{graphicx}	
\usepackage{amsmath}	
\usepackage{amssymb}	
\usepackage{subfigure}
\usepackage{threeparttable}
\usepackage{balance}
\usepackage{indentfirst}
\usepackage{CJKutf8}
\usepackage{siunitx}
\usepackage{array}
\usepackage{booktabs}
\usepackage{rotating}
\usepackage{mathrsfs}
\usepackage{xcolor}

\newcommand{\lv}{\ifmmode L_{5100} \else $L_{5100}$\fi}
\newcommand{\kms}{\ifmmode {\rm km\ s}^{-1} \else km s$^{-1}$\fi}
\newcommand{\ergs}{\ifmmode {\rm erg\ s}^{-1} \else erg s$^{-1}$\fi}
\newcommand{\ergc}{\ifmmode {\rm erg\ s^{-1}cm^{-2}} \else $\rm erg\ s^{-1} cm^{-2}$\fi}
\newcommand{\lb}{\ifmmode L_{\rm Bol} \else $L_{\rm Bol}$\fi}
\newcommand{\ledd}{\ifmmode L_{\rm Edd} \else $L_{\rm Edd}$\fi}
\newcommand{\hb}{\ifmmode H\beta \else H$\beta$\fi}
\newcommand{\ha}{\ifmmode H\alpha \else H$\alpha$ \fi}
\newcommand{\oiii}{\ifmmode \rm [O\ \textsc{iii}] \else $\rm [O\ \textsc{iii}]$\fi}
\newcommand{\nii}{\ifmmode \rm [N\ \textsc{ii}] \else $\rm [N\ \textsc{ii}]$ \fi}
\newcommand{\mgii}{\ifmmode \rm Mg\ \textsc{ii} \else $\rm Mg\ \textsc{ii}$\fi}
\newcommand{\civ}{\ifmmode \mathrm{C}\ \textsc{iv} \else $\mathrm{C}\ \textsc{iv}$\fi}
\newcommand{\ciii}{\ifmmode \rm C\ \textsc{iii]} \else $\rm C\ \textsc{iii]}$\fi}
\newcommand{\siiv}{\ifmmode \rm Si\ \textsc{iv} \else $\rm Si\ \textsc{iv}$\fi}
\newcommand{\siiii}{\ifmmode \rm Si\ \textsc{iii} \else $\rm Si\ \textsc{iii}$\fi}
\newcommand{\siii}{\ifmmode \rm Si\ \textsc{ii} \else $\rm Si\ \textsc{ii}$\fi}
\newcommand{\heii}{\ifmmode \rm He\ \textsc{ii} \else $\rm He\ \textsc{ii}$\fi}
\newcommand{\aliii}{\ifmmode \rm Al\ \textsc{iii} \else $\rm Al\ \textsc{iii}$\fi}
\newcommand{\niii}{\ifmmode \rm N\ \textsc{iii} \else $\rm N\ \textsc{iii}$\fi}
\newcommand{\mbh}{\ifmmode M_{\rm BH}  \else $M_{\rm BH}$\fi}
\newcommand{\niv}{\ifmmode \rm N\ \textsc {iv} \else $\rm N\ \textsc {iv}$\fi}
\newcommand{\msun}{M_{\odot}}
\newcommand{\rfe}{\ifmmode R_{\rm Fe} \else $R_{\rm Fe}$\fi}
\newcommand{\sst}{\ifmmode \sigma_{\rm \ast}\else $\sigma_{\rm \ast}$\fi}
\newcommand{\dhb}{\ifmmode D_{\rm H\beta} \else $D_{\rm H\beta}$ \fi}
\newcommand{\leddR}{\ifmmode L_{\rm Bol}/L_{\rm Edd} \else $L_{\rm Bol}/L_{\rm Edd}$\fi}
\newcommand{\feii}{\ifmmode \rm Fe\ \textsc{ii} \else $\rm Fe\ \textsc{ii}$\fi}
\newcommand{\mdot}{\ifmmode \dot{\mathscr{M}}  \else $\dot{\mathscr{M}}$\fi}
\newcommand{\rhb}{\ifmmode R_{\rm BLR}({\rm H\beta})  \else $R_{\rm BLR}({\rm H\beta})$ \fi}
\newcommand{\shb}{\ifmmode \sigma_{\rm H\beta} \else $\sigma_{\rm \hb}$\fi}
\newcommand{\RL}{\ifmmode R_{\rm BLR}({\rm H\beta}) - L_{\rm 5100} \else $R_{\rm BLR}({\rm H\beta}) - L_{\rm 5100}$\fi}
\newcommand{\ms}{\ifmmode M_{\rm BH}-\sigma_{\ast} \else $M_{\rm BH}-\sigma_{\ast}$\fi}
\newcommand{\bd}{\ifmmode \rm \ha/\hb\ \else $\rm \ha/\hb$\fi}

\begin{document}

\title{Outflows and Accretion in High-Redshift Quasars: Insights from Gemini GNIRS and SDSS Spectroscopy}

\author[0009-0004-4431-4535]{Xiang-Wei Ning}
\affiliation{School of Physics and Technology, Nanjing Normal University, Nanjing 210023, People's Republic of China}
\email{xwning@njnu.edu.cn}

\author{Chen-Sheng Gu}
\affiliation{School of Physics and Technology, Nanjing Normal University, Nanjing 210023, People's Republic of China}
\email{1034061360@qq.com}

\author[0000-0002-2121-8960]{Wei-Hao Bian}
\affiliation{School of Physics and Technology,  Nanjing Normal University, Nanjing 210023, People's Republic of China}
\email[show]{whbian@njnu.edu.cn}

\author[0009-0008-5295-3977]{Yang-Song Ma}
\affiliation{School of Physics and Technology,  Nanjing Normal University, Nanjing 210023, People's Republic of China}
\email{2244180800@qq.com}

\author{Wen-Jie Hu}
\affiliation{School of Physics and Technology, Nanjing Normal University, Nanjing 210023, People's Republic of China}
\email{3492445582@qq.com}

\author{Zi-Han Yuan}
\affiliation{School of Physics and Technology,  Nanjing Normal University, Nanjing 210023, People's Republic of China}
\email{3492445582@qq.com}

\author{Yu-Meng Guan}
\affiliation{School of Physics and Technology,  Nanjing Normal University, Nanjing 210023, People's Republic of China}
\email{2304174572@qq.com}

\begin{abstract}
For a sample of 260 high-redshift quasars ($1.5 < z < 3.6$) with optical spectra from the Sloan Digital Sky Survey (SDSS) and near-infrared spectra from the Gemini Near-Infrared Spectrograph (GNIRS), we perform uniform spectral fitting of the optical and ultraviolet spectra in the rest frame to measure the properties of emission lines such as \hb, \feii, and \civ, which are used to investigate the connection between outflows traced by the \civ\ emission line and supermassive black hole (SMBH) accretion physics. 
We find that, compared to low-$z$ SDSS quasars, our high-$z$ sample exhibits larger mean \hb\ line widths as measured by both the full width at half maximum ($\rm FWHM_{\hb}$) and the line dispersion (\shb), but smaller mean values of the broad \hb\ shape parameter $D_{\rm \hb} \equiv \rm FWHM_{\hb}/\shb$ and the relative optical \feii\ strength (\rfe).
It is found that optical \rfe\ is moderately anti-correlated with \oiii\ equivalent width and strongly correlated with the $\rm FWHM_{\hb}$-based dimensionless SMBH accretion rate (\mdot), suggesting that the optical \rfe\ serves as a tracer of \mdot, as found in low-$z$ quasars. Moderate positive correlations are found between the UV \civ\ emission-line blueshift and both optical \rfe\ and \mdot. 
These relations support radiation-pressure-driven disk-wind models in which \civ\ emission-line outflow properties are governed by $\rm{FWHM}_{\hb}$ and \mdot.
\end{abstract}

\section{Introduction} 
Broad emission lines are the most prominent spectral features of Type I active galactic nuclei (AGNs) and quasi-stellar objects (quasars) \citep[e.g.,][]{netzer2013agnbook}. These lines are widely accepted to arise from the photoionization of high-velocity clouds within the broad-line regions (BLRs), which are situated within the gravitational influence of the central supermassive black hole (SMBH). Ionizing photons from the accretion disk surrounding the SMBH irradiate these BLR clouds, and the subsequent recombination processes produce the observed broad emission lines. The properties of these lines, including their widths, strengths, and profiles, provide valuable insights into the physical processes occurring in the central regions of AGNs and quasars \citep[e.g.,][]{baskin2005,richards2011unification, netzer2013agnbook,wang2017}.

The blueshift of the \civ\ emission line relative to low-ionization lines is commonly found in quasar ultraviolet (UV) spectra \citep[e.g.,][]{gaskell1982redshift,richards2002broad, richards2011unification, ge2019blueshift, matthews2023gemini}. The accuracy of the \civ\ emission line blueshift depends on the reliability of the systemic redshift. Several methods are employed to determine the redshift of quasars. One primary approach involves measuring redshift from the absorption lines of the host galaxy. A secondary method makes use of the narrow \oiii\ \(\lambda\)5008 and Balmer lines, which generally exhibit smaller shifts compared to the host absorption lines. However, at higher redshift ($z>0.8$), these lines move into the near-infrared band, rendering them undetectable by optical telescopes. Another option for determining systemic redshift is to use the UV \mgii\ or high-ionization \civ\ emission line. Nevertheless, \civ\ often exhibits blueshift, which makes using the \civ\ emission line to determine redshift an unreliable method. 
Using rest-frame optical and UV spectral emission lines from the Sloan Digital Sky Survey (SDSS) reverberation-mapping (RM) sample to determine systemic redshift, the uncertainties in redshift determined from \oiii\ \(\lambda\)5008, \mgii, and \hb\ emission lines are approximately 50 \kms, 200 \kms, and 400 \kms, respectively \citep{shen2016sloan}.

Quasars often exhibit very strong and distinct broad absorption lines (BALs) in the UV spectra \citep[e.g.,][]{gibson2009catalog, lyke2020sloan, Rankine2020}. BAL quasars typically have absorption lines with widths of 2000 \kms\ and can accelerate to greater than 0.1 times the speed of light, usually identified from high-ionization UV lines such as \civ\ and \siiv. BALs are believed to originate from the absorption of radiation by high-speed outflowing gas that lies along the line of sight, accelerated at scale of tens of parsecs primarily by dust-driving \citep{He2022}, although radiation-line driving may dominate at smaller scales within BLRs \citep{Murray1995, Proga2000}. BAL quasars are considered as a special type of quasar, possibly in different evolutionary stages or due to different viewing angles \citep{trump2006catalog, gibson2009catalog}. The blueshift measurement of the \civ\ emission line for BAL quasars depends on the spectral decomposition including absorption lines \cite[e.g.,][]{ge2019blueshift}.

For a sample of 67 low-$z$ Palomar–Green (PG) quasars and assembled 112 high-$z$ quasars from different literatures with rest frame optical and UV spectra  ($0 < z < 5$), \cite{ge2019blueshift} performed a decomposition of the \civ\ $\lambda 1549$ broad-line profile \citep{ge2016}. Based on systemic redshift defined by \oiii\ $\lambda 5008$, the \civ\ emission line blueshift was found to have a moderate negative correlation with the \civ\ equivalent width (EW), which is consistent with the result from SDSS quasars by \cite{richards2011unification}. \cite{ge2019blueshift} also found a weak positive correlation between the \civ\ blueshift and the continuum luminosity at 5100 \AA\ (\lv) or the Eddington ratio (\leddR), where $L_{\rm Bol}$ is the bolometric luminosity and $L_{\rm Edd}$ is the Eddington luminosity.

Regarding the origin of the \civ\ blueshift, \cite{richards2011unification} proposed a two-component disk-wind model for the BLRs, based on two observed anti-correlations: between \civ\ EW and the UV continuum luminosity at 1350 \AA\ \citep[the Badwin effect, et al.,][]{Baldwin1977, ge2016}, and between \civ\ EW and the \civ\ emission-line blueshift. In this model, the BLR is not a static structure but consists of an outflowing wind launched from the accretion disk surrounding the central SMBH. This wind is primarily driven by radiation pressure, especially from UV photons that interact with ions in the gas through line scattering. The strength and structure of the wind depend on the spectral energy distribution (SED) of the ionizing radiation \citep[e.g.,][]{Rankine2020, Ma2026}. A stronger UV flux tends to enhance the wind by increasing radiation pressure. 
Conversely, a relatively strong X-ray flux can over-ionize the gas, reducing the number of effective scatterers and thereby suppressing the wind \citep{Leighly2004, Casebeer2006}. 
A stronger UV flux also tends to imply a softer ionized SED \citep[e.g.,][more negative $\alpha_{\rm ox}$]{Liu2021}, leading to smaller \civ\ EW, although this has recently been challenged by \cite{Cai2023}. As a result, the disk-wind model  can explain \civ\ outflow and its relation with the \civ\ EW, both of which are governed by the SED.
The connection between wind properties and the underlying accretion process, mediated by the SED, can give clues for interpreting the complex spectral behavior of AGN. 
Recent hydrodynamic simulations further demonstrate that the coexistence of jets and winds plays a crucial role in AGN feedback efficiency \citep{He2025}.

For the underlying accretion process in AGNs, there are some key parameters, such as the SMBH mass (\mbh), the SMBH spin, and the accretion rate ($\dot{M}$). For Type I AGNs, the BLR clouds emitting the broad \hb\ line can be used as a probe to calculate the virial mass  \citep[e.g.,][]{Peterson04, netzer2013agnbook}:
 \begin{equation}
 \label{eq1}
\mbh=f\times \frac{R_{\rm BLR}~(\Delta V)^2}{G}  .
\end{equation}
where $\Delta V$ is the velocity of the BLR clouds traced by the the full width at half maximum ($\rm FWHM_{\hb}$) or the line dispersion (\shb), $f$ is a corresponding virial factor \citep[][and references therein]{Ho2014, yu2020supermassive, Li2025}, and $G$ is the gravitational constant. The BLRs size $R_{\rm BLR}$ can be measured by AGNs RM method.
For other un-RM AGNs, an empirical $R_{\rm BLR}-\lv$ relation is used to estimate $R_{\rm BLR}$ from \lv\ \cite[e.g.,][]{kaspi2000reverberation, vestergaard2006determining, shen2011dr7,yu2020supermassive}.  
Through investigation of super-Eddington accreting massive black holes \cite[SEAMBHs,][]{wang2013super}, it was found that the accretion rate has an influence on the empirical $R_{\rm BLR}-\lv$ relation \citep[also see][]{Wu2026}.
Considering that \rfe\ is correlated with Eigenvector 1 and is shown to be driven by the Eddington ratio \citep{boroson1992green,shen2014nature, ge2016}, \rfe\ can be used as a proxy for the accretion rate, leading to an extended empirical $R_{\rm BLR}-\lv$ relation \citep{du2019radius,yu2020extendedRL}:
\begin{equation}\label{eq2}
	\begin{aligned}
		\log \frac{R_{\mathrm{BLR}}(\mathrm{H} \beta)}{\text{lt-day}} &= (0.48 \pm 0.03) \log l_{\rm 44}  \\
		 &- (0.38 \pm 0.04) R_{\mathrm{Fe}} + (1.67 \pm 0.09),
	\end{aligned}
\end{equation}
where $l_{44} = \lv/10^{44}\ergs$ .

For the high-ionization lines, such as \civ, it was found that the \civ-based \mbh\ is biased with respect to the \hb-based \mbh\ \citep[e.g.][]{baskin2005, bian2012, shen2012, Brotherton2015,coatman2017, Dix2023}. The uncertainties in the determination of \mbh\ may arise from the non-virial component shown in the line profile, such as from a radiation-driven disk wind \cite[e.g.,][]{Murray1995, richards2011unification}. In order to investigate wind properties through the \civ\ blueshift with the underlying accretion process, we need the infrared spectra of high-$z$ quasars to estimate their \hb-based \mbh\ and subsequently derive the accretion rates.

In our previous study of \civ\ emission-line blueshift \citep{ge2019blueshift}, we used a small sample of QSOs with infrared spectra spanning wide ranges of source-selection criteria, instrument properties, spectral band and resolution, and signal-to-noise ratio (S/N).
In this paper, we present uniform spectral decomposition for a sample of 260 high-$z$ quasars with available near-infrared (NIR) spectra from Gemini Near-infrared Spectrograph (GNIRS) and optical SDSS spectra \citep{matthews2023gemini}, and investigate the \civ\ emission-line blueshift relative to systemic redshifts measured from the line centres of \oiii, \mgii, or \hb, as well as its connection to the SMBH accrection physics through  with \hb-based \mbh\ estimates. The sample and analysis are described in Section \ref{s2}. Our results and discussions are given in Section \ref{s3}. Finally, our conclusions are presented in Section \ref{s4}. All of the cosmological calculations in this paper assume \(\Omega_{\Lambda} = 0.7\), \(\Omega_{\rm M} = 0.3\), and \(H_{0} = 70 \, \text{km} \, \text{s}^{-1} \, \text{Mpc}^{-1}\). 

\section{Samples and Spectral Decomposition} \label{s2}
\subsection{Samples}
\begin{figure}
    \centering
        \includegraphics[width =3.5in]{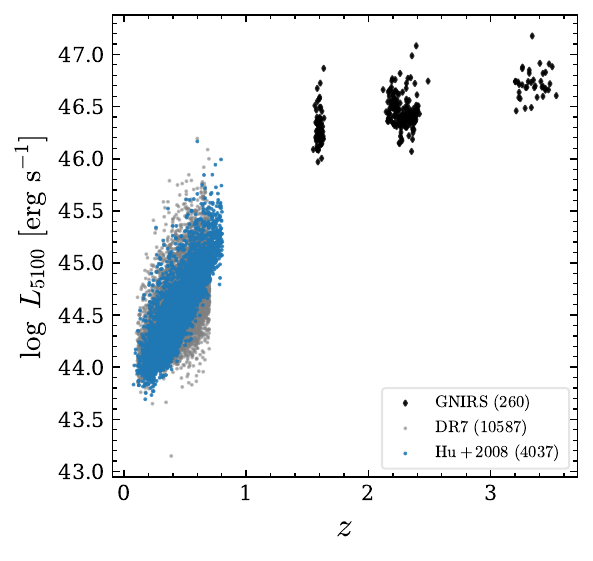}
    \caption{Rest-frame 5100 \AA\ luminosity ($L_{5100}$) versus redshift ($z$) for different samples: GNIRS (black rhombuses), SDSS DR7 (gray circles), \citet{hu2008systematic} (blue circles).}
    \label{fig1}
\end{figure}

Utilizing GNIRS at the Gemini North Observatory,  \cite{matthews2021gnirs} observed a flux-limited ($m_{\rm i}<19.0$, $ H<16.5$) sample of quasars with $1.5<z<3.6$ in the $0.8$--$2.5~ \mu\rm m$ band (Gemini Near Infrared Spectrograph-Distant Quasar Survey, GNIRS-DQS). The sources were selected from the quasars catalogues of SDSS DR12 and DR14 \citep{paris2017sloan,paris2018sloan}. The three redshift ranges for the observed quasars are $1.55 \leq z \leq 1.65$, $2.10 \leq z \leq 2.40$, and $3.20 \leq z \leq 3.60$, which place the \hb\ region in the highest S/N regions of the $J$, $H$, and $K$ bands, with a spectral resolution of $R\sim 1100$.
Sources in the redshift range of 2.10 to 2.40 account for $67\%$ of the total sample. 
At last a sample of 260 high-$z$ SDSS quasars with near-infrared spectra was provided by \cite{matthews2021gnirs, matthews2023gemini}
\footnote{The sample used in this study is available at \url{https://datalab.noirlab.edu/gnirs_dqs.php}.}
 (Figure \ref{fig1}). 
The systemic redshifts used in the fitting below were provided by \cite{matthews2023gemini}, measured from the line centres of \oiii, \mgii, or \hb. Among them, systemic redshifts for 182 sources were determined from \oiii.

We also select some samples of low-$z$ quasars shown in the plane of $L_{\rm 5100}$ versus $z$ (see Figure \ref{fig1}). 
\cite{hu2008systematic} selected 4037 low-$z$ quasars with $z < 0.8$ from the SDSS DR5 quasar catalogue \citep{schneider2007sloan}. The SDSS spectral resolution $R \approx 1850-2200$, with a wavelength interval in log space of $10^{-4} $, corresponding to a velocity of $\sim 69~\kms$. 
Specific selection criteria includes S/N greater than 10 in the rest-frame wavelength range of 4430-5550 \AA. The sources with following features are excluded: the reduced $\chi^2$ in continuum fitting is greater than 4; optical \feii\ equivalent width is less than 25 \AA;  \hb\ broad component FWHM error is greater than 10\%; \oiii\ $\lambda$ 5008 peak velocity shift error is greater than 100 \kms.
A catalog of 105,783 quasars from SDSS DR7 is provided by \cite{shen2011dr7}. The DR7 catalog provides properties of the broad \ha, \hb, \mgii, and \civ\ emission lines. \cite{shen2014nature} used this quasar catalogue to study the diversity of quasars. 

\subsection{Spectral Decomposition}

\begin{figure*}
    \centering
        \includegraphics[width = 4in]{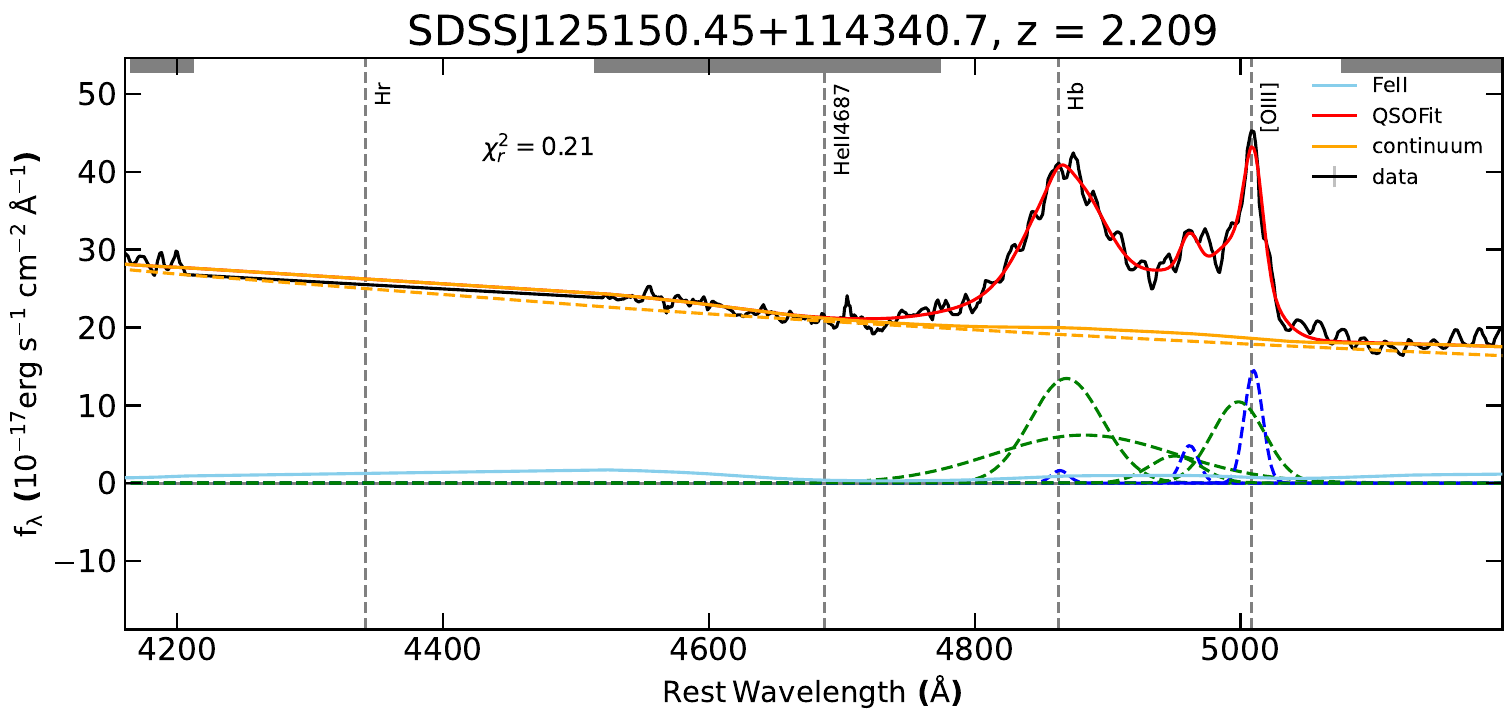}
        \includegraphics[width = 2.45in]{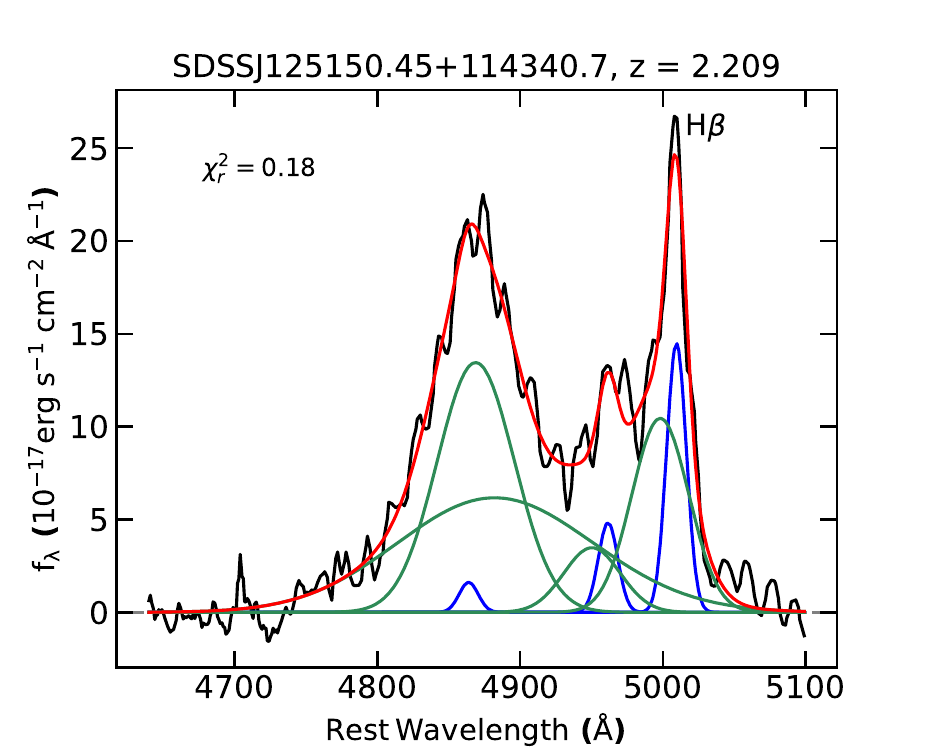}
        \hspace*{-0.1in}
        \raisebox{0.04in}{\includegraphics[width = 4.05in]{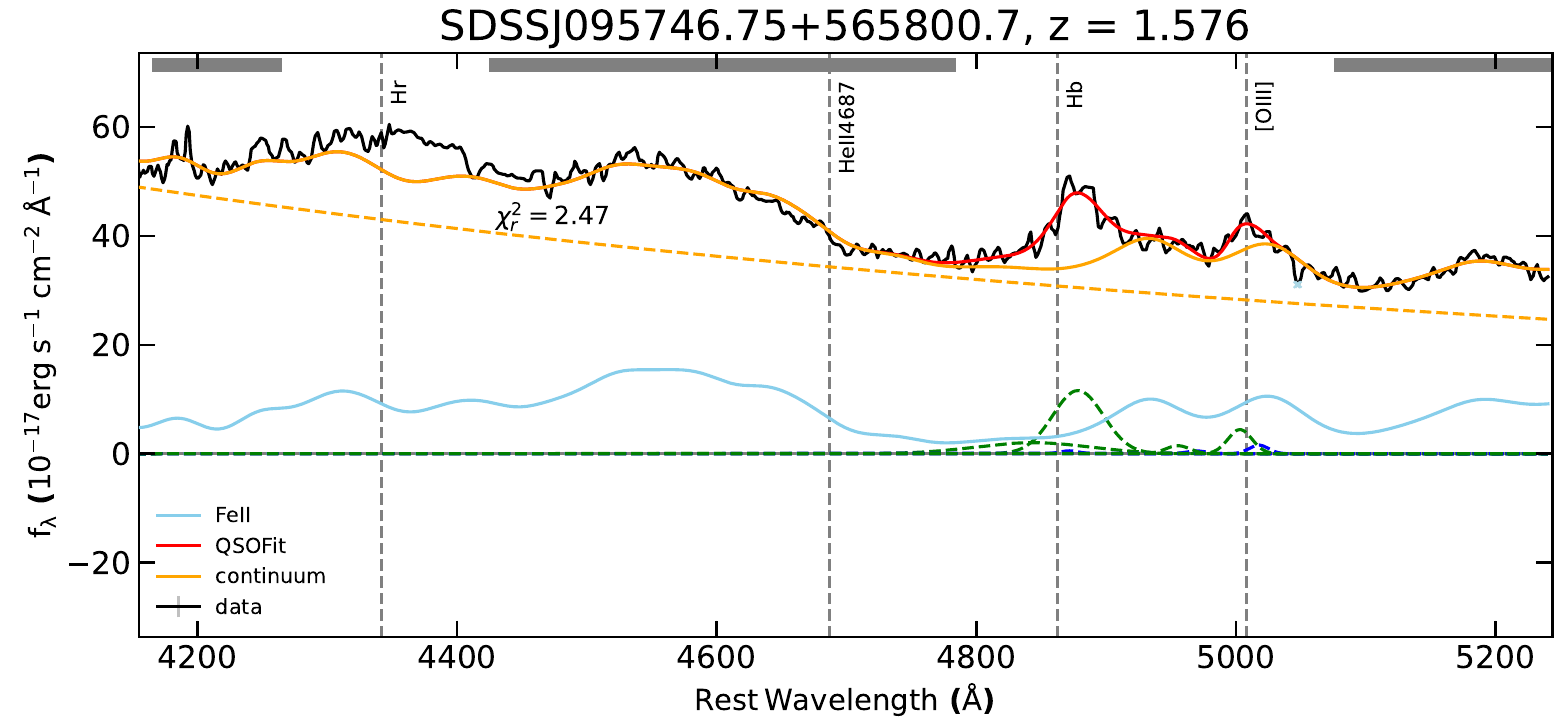}}
        \includegraphics[width = 2.42in]{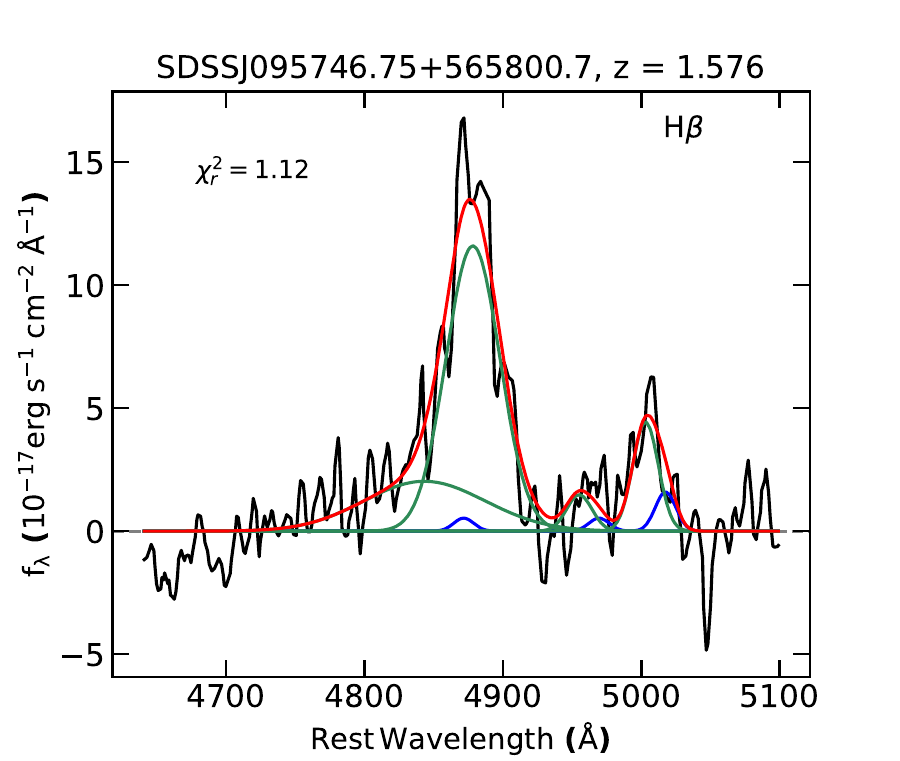}
    \caption{Two examples of the GNIRS spectral decomposition: SDSS J125150.45+114340.7 with weak \rfe\ (top) and SDSS J095746.75+565800.7 with strong \rfe\ (bottom). Left: global fitting (pseudo-continuum and line fitting). The black solid line shows the spectral data; the orange solid line shows the pseudo-continuum which includes power-law continuum (orange dashed) and optical \feii\ (light blue). Also shown are the broad Gaussian component (green dashed), the narrow Gaussian component (blue dashed), and total fitted profile (red solid). The gray dashed lines show the wavelengths of H$\gamma$, \heii, \hb, and \oiii\ emission lines, respectively. Right: line fitting around \hb. The broad (green solid) and narrow (blue solid) Gaussian component are shown, along with the total fitted profile (red solid). Reduced $\chi^2$ values are given in each panel.}
    \label{fig2}
\end{figure*}

\begin{figure*}
    \centering
        \includegraphics[width = 3in]{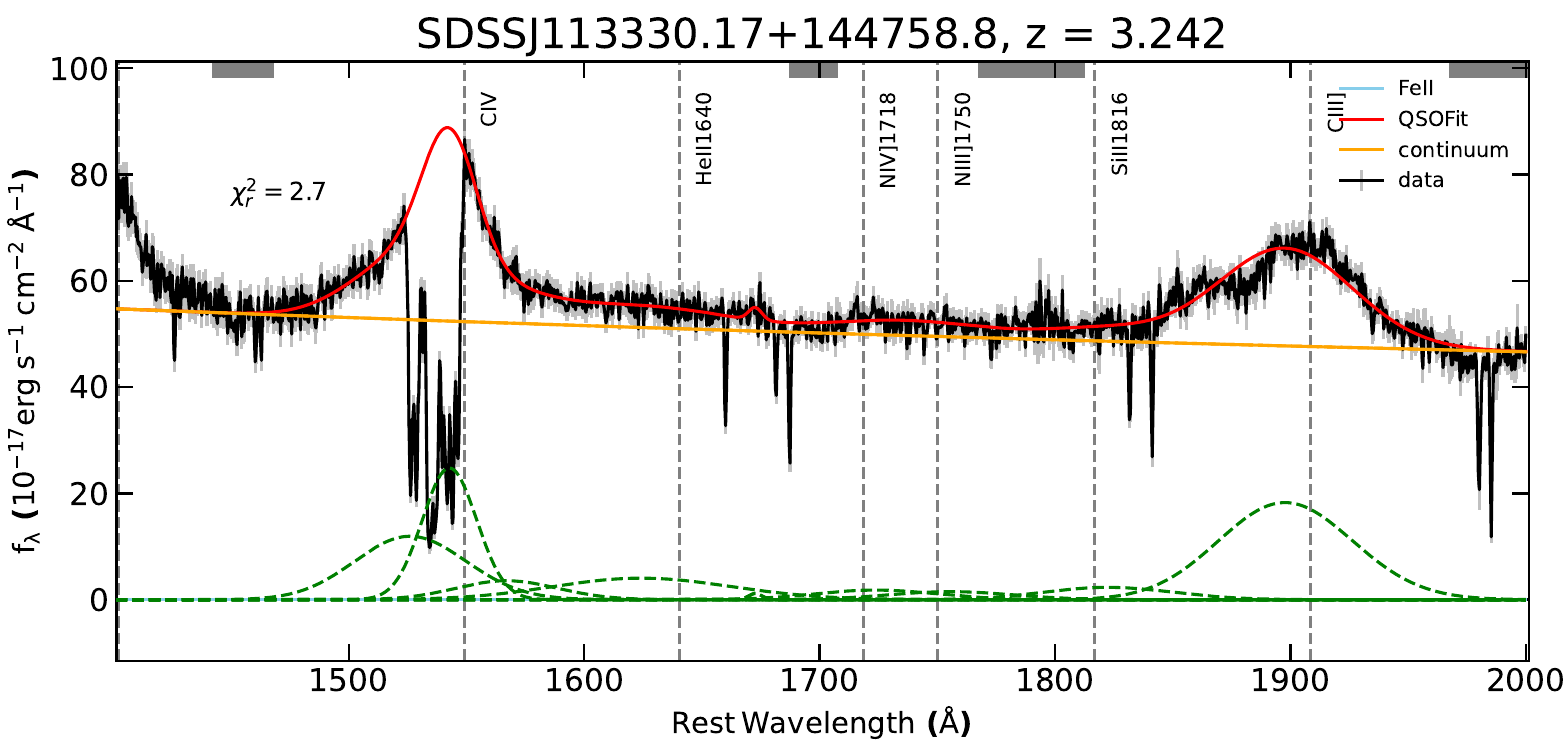}
        \includegraphics[width = 3.12in]{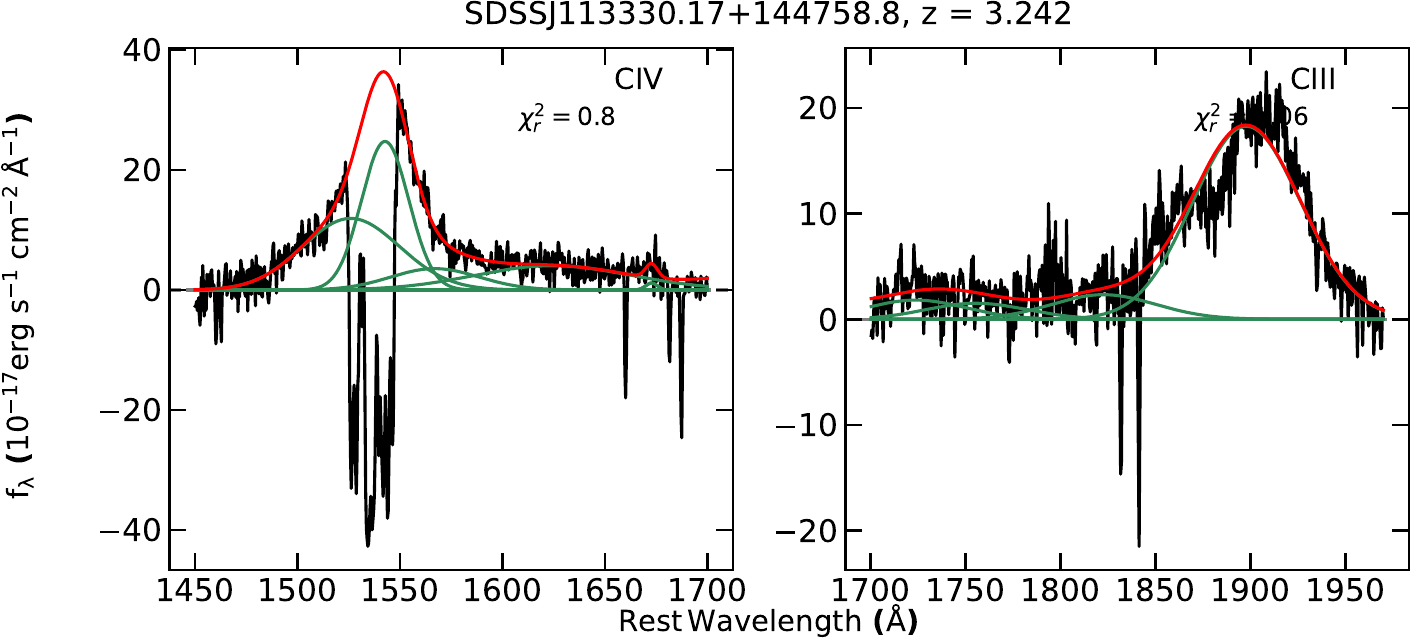}
    \caption{An example of the SDSS spectral decomposition for the BAL quasars SDSS J113330.17+144758.8. Left: global fitting (pseudo-continuum and lines fitting). Right: line fitting around the \civ\  and \ciii\ regions. Symbols and lines are the same as in Figure \ref{fig2}.}
    \label{fig3}
\end{figure*}

For the 260 high-$z$ quasars from GNIRS-DQS, the near-infrared spectra from GNIRS and the optical spectra from SDSS are fitted using PyQSOFit \citep{guo2018pyqsofit, Shen2019, wu2022dr16}. The GNIRS spectra have already been corrected for Galactic extinction; here we apply only the redshift correction to the rest frame, adopting the systemic redshifts from \cite{matthews2023gemini}. The spectra are smoothed with a three-pixel boxcar (see Figure~\ref{fig2}). 

About the near-infrared spectra from GNIRS, the average S/N in the rest-frame wavelength range of 5000--5200 \AA\ is measured to calculate the errors. The spectral fitting range is approximately 2900 -- 5600 \AA\ in the rest frame. We adjusted this range manually for each object (e.g., to avoid the low S/N near the edges of the spectra), and the adopted rest-frame fitting range is given in Table~\ref{table1}. The pseudo-continuum consists of a power-law spectrum and optical \feii. The optical \feii\ template from \cite{boroson1992green} is convolved with a Gaussian function, with all three parameters left free. The \feii\ FWHM range is set to 900 -- 18000 \kms, and the \feii\ shift range to $\pm 3000~ \kms$ relative to the systemic redshift. The pseudo-continuum fitting windows are: 2925 -- 3400, 3520 -- 3699, 3996 -- 4080, 4170 -- 4260, 4430 -- 4780, 5080 -- 5535. \(\chi^{2}\) minimum within these windows is performed using $lmfit$, with iterative 3\(\sigma\) clipping. GNIRS spectra are affected by telluric absorption; therefore, data in the spectral observed bands of 1.35 -- 1.45 \(\mu\)m and 1.8 -- 1.92 \(\mu\)m are excluded during the pseudo-continuum fitting \citep{burke2024absor}. Figure \ref{fig2} shows two samples of the spectral decomposition for SDSS J125150.45+114340.7 with weak \rfe\ (top) and SDSS J095746.75+565800.7 with strong \rfe\ (bottom).

About the optical emission lines around \hb, the emission lines of \hb\ and \oiii\ are decomposed into multiple components. For \oiii\ \(\lambda\) 5008, two Gaussian functions are used for fitting: one Gaussian set as the narrow component representing the core component, another Gaussian set as the wing component.
Then, keeping the line widths and the shifts relative to the rest-frame line centres of these two Gaussians unchanged, they are shifted and scaled to fit \oiii\ \(\lambda\)4960, with the flux ratio set to 3 for the corresponding Gaussians (core, wing). For the narrow component contribution of \hb, considering the quality of the spectral data, a narrow Gaussian is used for fitting, with its line width and shift relative to the line center tied to the core component of \oiii. The flux ratio of the \hb\ narrow component to \oiii\ \(\lambda\)5008 is set to no more than 1/3. For the broad \hb\ emission line, two broad Gaussians are used for fitting. The broad Gaussian velocity field is set to be greater than or equal to 1200 \kms, and the narrow Gaussian is set to be less than 1200 \kms. The \(\chi^{2}\) minimum fitting for the emission lines is performed using $lmfit$. 
The errors of the fitting parameters are given by Markov Chain Monte Carlo (MCMC) simulations.

At the same time, the SDSS spectra of these 260 high-$z$ quasars (covering the \civ\ emission line in the rest frame) are retrieved from SDSS website \footnote{\url{https://www.sdss.org/}}.
Galactic extinction correction is applied using the Galactic dust distribution map from \cite{schlegel1998maps} and the Galactic extinction curve from \cite{cardelli1989relationship} (\(R_{\rm V} = 3.1\)). The systemic redshift provided by \cite{matthews2023gemini} is used for redshift correction.
Due to the complexity of the ultraviolet spectra, a segmented fitting method is adopted. 
The main emission lines \civ\ + \ciii\ were fitted in the rest-frame wavelength range of 1400 -- 2000 \AA\ for all objects. For the \civ+ \ciii\ regions, the pseudo-continuum includes a single power-law and UV \feii\ multiples \citep[e.g.,][]{Shen2019, Zhao2020, Pan2025}. The UV \ion{Fe}{2} template in PyQSOFit combines the \ion{Fe}{2} template from \cite{vestergaard2001empirical}, \cite{Salviander2007}, and \cite{Tsuzuki2006} for the wavelength ranges 1000 -- 2200, 2200 -- 3090, and 3090 -- 3500~\AA, respectively. The fitting windows for the pseudo-continuum are: 1445 -- 1465, 1690 -- 1705, 1770 -- 1810, and 1970 -- 2000 \AA. 

Emission line fitting is then performed on the pseudo-continuum subtracted spectra.
Considering the blueshift and broad absorption lines present in the \civ\ emission line, three broad Gaussian functions are used to fit the \civ\ emission line components. 
\heii\ \(\lambda\)1640 and O\text{\small\ III} \(\lambda\)1663 are each fitted with one broad and one narrow Gaussian. For the \ciii\ emission line region, \ciii\ is fitted with two broad Gaussian functions, and other emission line components in this region, such as \siiii\ \(\lambda\)1892,
 \aliii\ \(\lambda\)1857, \siii\ \(\lambda\)1816, \niii\ \(\lambda\)1750, and \niv\ \(\lambda\)1718, are each fitted with one broad Gaussian.
Iterative refitting is performed for the emission lines. Due to the influence of \civ\ broad absorption lines, some sources still have incomplete emission line fitting. The \civ\ and \ciii\ regions are visually inspected, and broad absorption troughs are manually masked to obtain reasonable \civ\ emission line profile fits. In general, although most BAL quasars share the same continuum windows as non-BAL quasars, we manually adjusted the continuum windows for some complex BAL quasars whose absorption troughs extend into the continuum windows. Figure \ref{fig3} shows a fitting example of a BAL quasars. Rest-frame UV and optical spectral properties for 260 high-$z$ quasars are shown in Table~\ref{table1}.

It is worth mentioning that our spectral fitting differs from that of \cite{matthews2023gemini} and \cite{wu2022dr16} in several aspects. For the GNIRS spectra, we model the AGN continuum as a power-law component plus \ion{Fe}{2} emission, whereas \cite{matthews2023gemini} used localized linear continua. The median difference between our measurements and those of \cite{matthews2023gemini} for \rfe, $\mathrm{log}~\lv$, $\rm log~FWHM_{\hb}$, and logarithmic [\ion{O}{3}]$~\lambda5008$ FWHM are $0.07\pm0.30$, $-0.03\pm0.03$, $0.06\pm0.07$, and $0.02\pm0.23$, respectively. For the SDSS spectra, we perform local spectral fitting (\citealt{wu2022dr16} perform global spectra fitting), our AGN continuum includes only a power-law component and \ion{Fe}{2} emission (\citealp{wu2022dr16} additionally include a third-order polynomial), and we carefully masked the \ion{C}{4} absorption troughs (\citealp{wu2022dr16} do not). The median differences between our measurements and those of \cite{wu2022dr16} for \ion{C}{4} peak wavelength, logarithmic \ion{C}{4} FWHM, and logarithmic continuum luminosity at 1700 \AA\ are $-0.51\pm5.03$, $0.01\pm0.15$, and $0.01\pm0.04$, respectively. Overall, our measurements well agree with those of \cite{matthews2023gemini} and \cite{wu2022dr16}.

\section{Results and Discussions}\label{s3}
\subsection{The optical spectral properties in the rest frame for the high-$z$ quasars}

\begin{figure*}
	\centering
			\includegraphics[angle=0,width=2.5in]{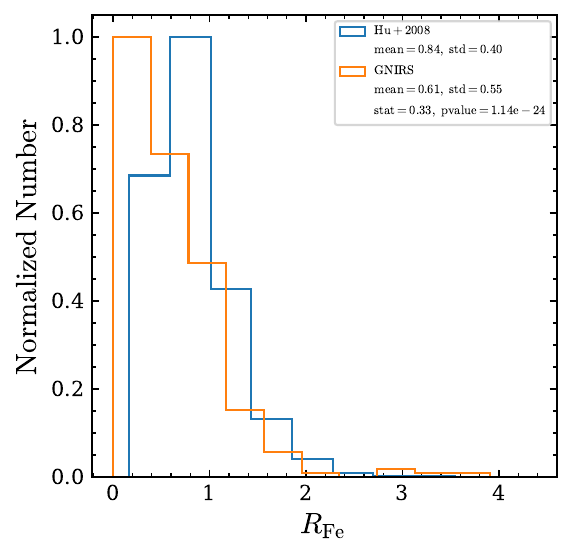}
	        \includegraphics[angle=0,width=2.5in]{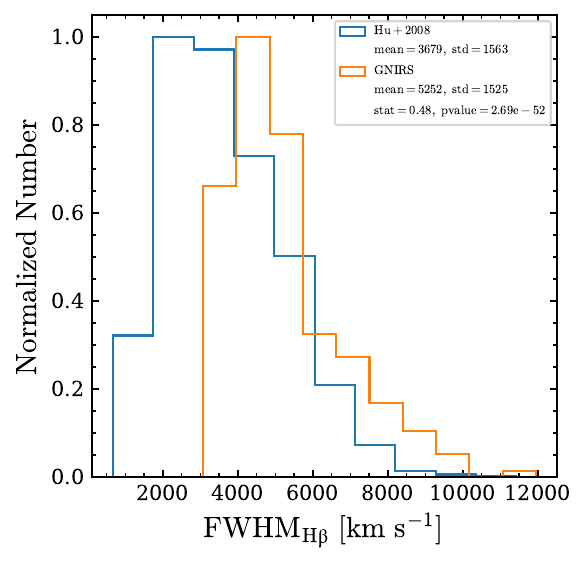}
			\includegraphics[angle=0,width=2.5in]{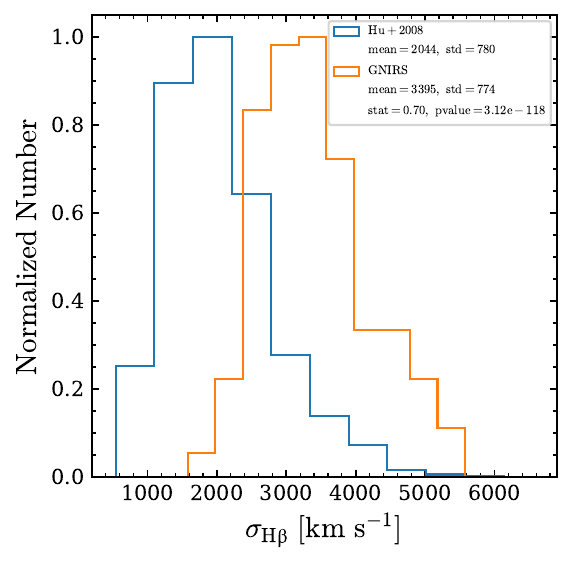}
			\includegraphics[angle=0,width=2.5in]{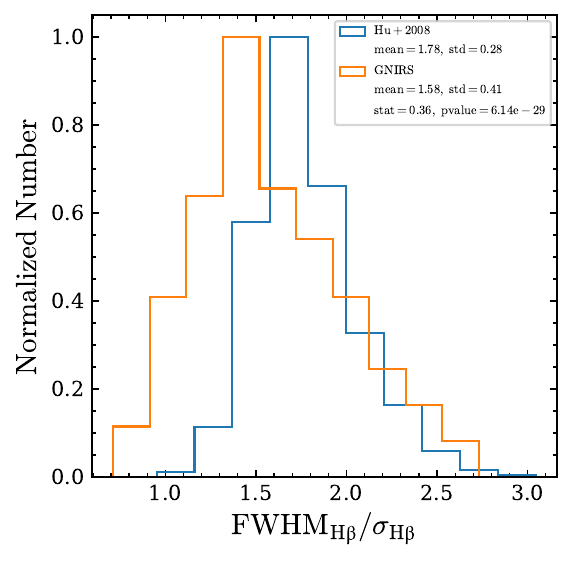}
	\caption{Distributions of \rfe, $\rm FWHM_{\hb}$, $\sigma_{\rm \hb}$, and \dhb for our high-$z$ sample (orange) and the low-$z$ sample of \cite{hu2008systematic} (blue), arranged from left to right and top to bottom. 
}
	\label{fig4}
\end{figure*}

\begin{figure*}
	\centering
			\includegraphics[angle=0,width=3.4in]{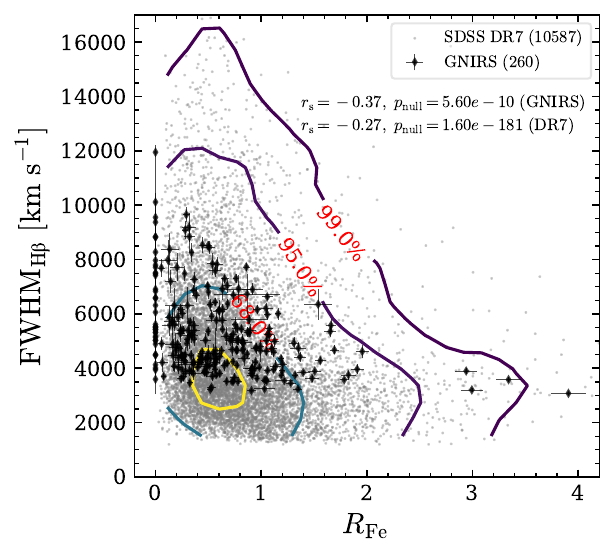}
	        \includegraphics[angle=0,width=3.2in]{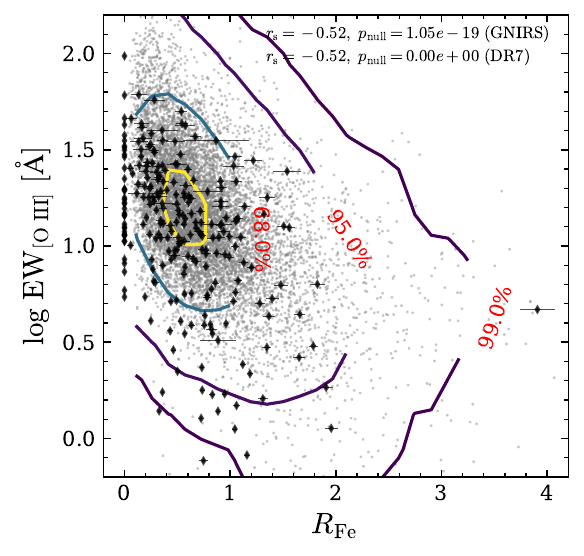}
	\caption{Left: $\rm FWHM_{\hb}$ versus \rfe. Right: \oiii\ $\lambda 5008$ EW versus \rfe. Gray points shows the SDSS DR7 sample; black rhombuses, our high-$z$ sample. Contours describe the density of DR7 points.}
	\label{fig5}
\end{figure*}

For the broad \hb\ emission line, $\rm FWHM_{\rm \hb}$, \shb, EW, integrated flux, and the line profile shape $D_{\rm \hb}=\rm FWHM_{\rm \hb}/\shb$ are calculated from the composite profile of all the broad \hb\ components \citep{boroson1992green}. 
The relative strength of \feii, $\rfe\ \equiv \frac{f_{\feii}}{f_{\rm \hb}}$, is defined as the ratio of the integrated flux of the \feii\  emission line in the 4434-4484 \AA\ range ($f_{\feii}$) to the integrated flux of the broad \hb\ component ($f_{\rm \hb}$), and the \rfe\ error is derived from their errors ($f_{\feii}^{\rm err}, f_{\rm \hb}^{\rm err}$) through error propagation:
\begin{equation}\label{eq3}
	\delta_{\rfe} = \sqrt{\left(\frac{f_{\feii}^{\rm err}}{f_{\feii}}\right)^2 + \left(\frac{f_{\rm \hb} ^{\rm err}}{f_{\rm \hb}}\right)^2} \times \rfe
\end{equation}

Figure \ref{fig4} shows, from left to right and top to bottom, the distributions of \rfe, $\rm FWHM_{\hb}$, $\sigma_{\rm \hb}$, and \dhb for  the high-$z$ sample (orange), and the low-$z$ sample of \cite{hu2008systematic} (blue). 
The high-$z$ sample has a mean \rfe\ of \(0.609 \pm 0.55\), smaller than \(0.837 \pm 0.40\) for the low-$z$ sample. 
The mean \(\rm FWHM_{\rm \hb}\) and \shb\ for the high-$z$ sample are \(5252 \pm 1525\) \kms\ and \(3395 \pm 774\) \kms, respectively, larger than \(3679 \pm 1563\) \kms\ and \(2044 \pm 780\) \kms\ for the low-$z$ SDSS sample.
For the high-$z$ sample, $\dhb$ is characterized by larger $\rm FWHM_{\hb}$ and \shb\ but a smaller \(D_{\rm \hb}\). 
This is because the \hb\ emission line in the high-$z$ sample tends to have a broader wing component, which contributes more to \shb\ than to \(\rm FWHM_{\rm \hb}\). A two-sample KS test was performed on these four distributions (\rfe, \(\rm FWHM_{\rm \hb}\), \(\sigma_{\rm \hb}\), \(D_{\rm \hb}\)) \footnote{ \url{https://docs.scipy.org/doc/scipy/reference/generated/scipy.stats.ks_2samp.html} }. 
The p-values are all less than \(10^{-24}\), indicating that the two samples are draw from different underlying distributions.

Left panel in Figure \ref{fig5} is $\rm{FWHM}_{\hb}$ versus \rfe, known as the eigenvector 1 plane. The gray background points are 10587 low-$z$ quasars from the SDSS DR7 ($z \leq 0.7$), the black rhombuses are from our high-$z$ quasar sample. The contours describe the density of the SDSS DR7 sample points. With relative to the low-$z$ sample, the mean value of \rfe\ tends to be smaller and $\rm FWHM_{\hb}$ larger for high-$z$ sample. 
Right panel in Figure \ref{fig5} shows the relationship between \rfe\ and the \oiii\ \(\lambda\)5008 EW. As \rfe\ increases, \oiii\ EW tends to decrease \citep{shen2014nature}. The Spearman correlation coefficient \(r_{\rm s} = -0.52\) and the null hypothesis probability \(p_{\rm null} = 1.05 \times 10^{-19}\) indicate a significant moderate anti-correlation, which is consistent with the low-$z$ SDSS quasars from DR7 ($r_{\rm s} = -0.52$).

\subsection{$\rm \hb$-based \mbh\ and \mdot}

\begin{figure*}
	\centering
			\includegraphics[angle=0,width=2.5in]{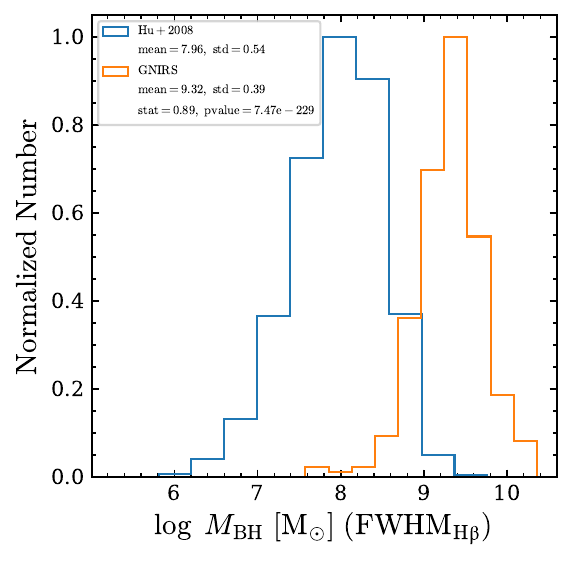}
	        \includegraphics[angle=0,width=2.5in]{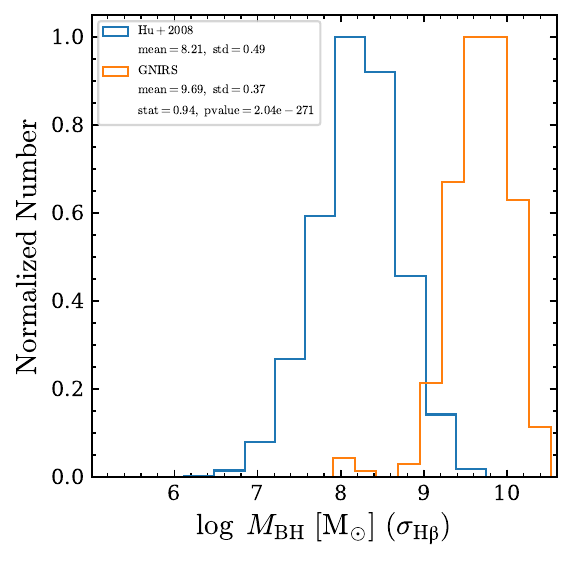}
			\includegraphics[angle=0,width=2.5in]{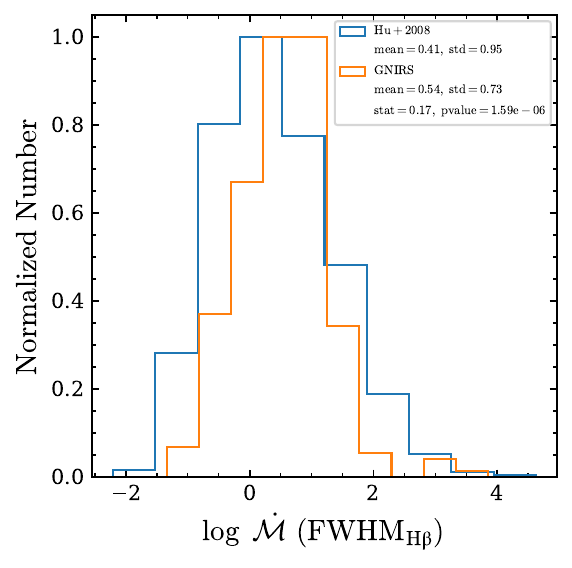}
			\includegraphics[angle=0,width=2.5in]{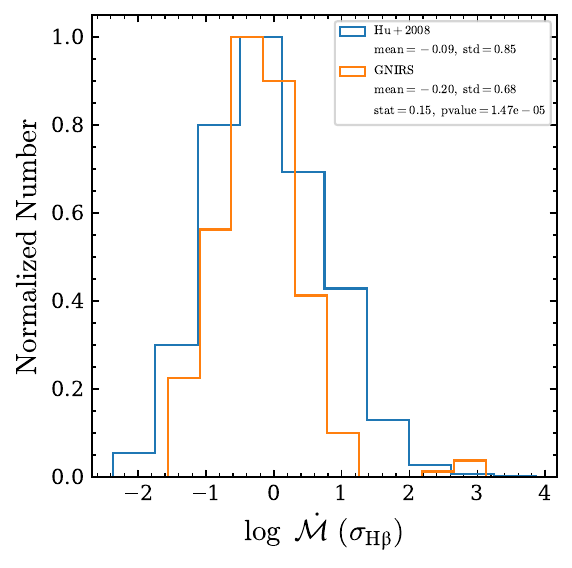}
			\caption{Distributions of $\rm FWHM_{\hb}$-based \mbh, $\sigma_{\rm \hb}$-based \mbh, $\rm FWHM_{\hb}$-based  \mdot, and $\sigma_{\rm \hb}$-based \mdot\ for our high-$z$ sample (orange) and the low-$z$ sample of \cite{hu2008systematic} (blue), arranged from left to right and top to bottom.
			}
	\label{fig6}
\end{figure*}

\begin{figure*}
	\centering
	\includegraphics[width=3.2in]{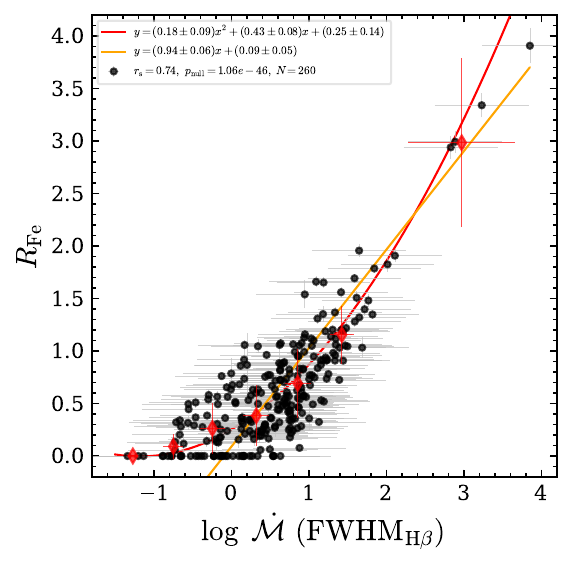}
	\includegraphics[width=3.2in]{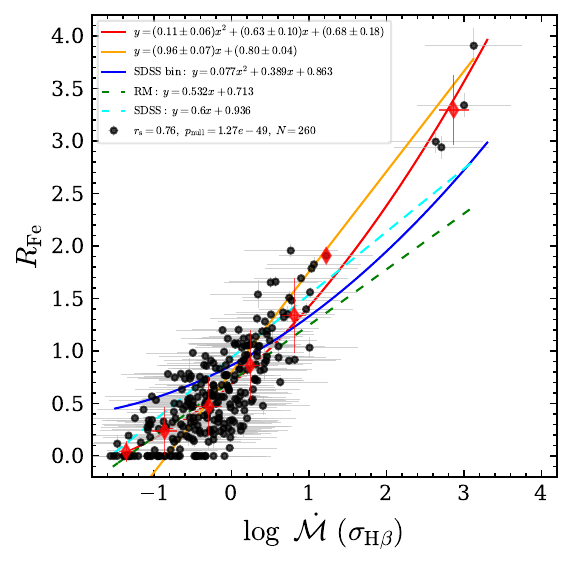}
	\caption{Left: \rfe\ versus $\rm FWHM_{\hb}$-based \mdot. The orange line shows the best linear fit with $r_{\rm s}$ and $p_{\rm null}$ in the bottom right corner and the RMS scatter in the upper left. The red line shows second-order polynomial fit to data binned in 0.6 dex intervals of \mdot. Right: \rfe\ versus $\sigma_{\rm \hb}$-based \mdot. The orange and red lines are as in the left panel. The green dashed line shows the best linear fit for RM AGNs \citep{liu2022bian}. The cyan dashed line shows the best linear fit for SDSS quasars, and the blue line shows the second-order polynomial fit \citep{liu2022bian}.}
		\label{fig7}
\end{figure*}

From the \rfe-corrected \hb\ broad-line region size formula (Equation \ref{eq2}), adopting $\rm FWHM_{\hb}$ and $f=1$ \citep{du2019radius},   $\rm FWHM_{\hb}$-based \mbh\ is derived as follows \citep{liu2022bian}:
\begin{equation}\label{eq4}
	\begin{aligned}
		 \log \frac{M_{\mathrm{BH}, \mathrm{FWHM}_{\rm H \beta}}}{M_{\odot}} &= 6.96 + 2 \log \frac{\mathrm{FWHM}_{\rm H \beta}}{1000 \mathrm{~km} \mathrm{~s}^{-1}}    \\
	& + (0.48 \pm 0.03) \log l_{44} - (0.38 \pm 0.04) R_{\mathrm{Fe}}
	\end{aligned}
\end{equation}
Adopting \shb\ and $f=5.5$ \citep{yu2020supermassive, yu2020extendedRL}, $\shb$-based \mbh\ is derived as follows \citep{liu2022bian}:
\begin{equation}\label{eq5}
	\begin{aligned}
		\log \frac{M_{\mathrm{BH}, \sigma_{\mathrm{H} \beta}}}{M_{\odot}} &= 7.7 + 2 \log \frac{\sigma_{\mathrm{H} \beta}}{1000 \mathrm{~km} \mathrm{~s}^{-1}}   \\
		 &+ (0.48 \pm 0.03) \log l_{44} - (0.38 \pm 0.04) R_{\mathrm{Fe}}
	\end{aligned}
\end{equation}

For the standard disk model with the effective disk temperature of $T_{\rm eff} \propto R^{-3/4}$, \mdot\ is \citep[e.g.,][]{SS73, Du2016}
 \begin{equation}
 \label{eq7}
\mdot\ \equiv \dot{M}/\dot{M}_{\rm Edd} = 20.1\left(\frac{l_{44}}{\cos\,\it{i}}\right)^{3/2} m_{7}^{-2}.
\end{equation}
where $m_{7}=M_{\rm BH}/10^7 \msun$, and ${\rm cos}~ i = 0.75$ is adopted \citep[e.g.,][]{liu2022bian}.

The dimensionless accretion rate \( \dot{\mathscr{M}} \) formulas calculated from two velocity tracers are:
\begin{equation}\label{eq8}
	\begin{aligned}
		\log \left(\dot{\mathscr{M}}_{\rm FWHM_{\mathrm{H} \beta}}\right) = & (0.54 \pm 0.06) \log l_{44} - 4 \log \frac{\mathrm{FWHM}_{\rm H \beta}}{1000 \mathrm{~km} \mathrm{~s}^{-1}}  \\
		& + (0.76 \pm 0.08) R_{\mathrm{Fe}} + 1.57
	\end{aligned}
\end{equation}
\begin{equation}\label{eq9}
	\begin{aligned}
		\log \left(\dot{\mathscr{M}}_{\sigma_{\mathrm{H} \beta}}\right) = & (0.54 \pm 0.06) \log l_{44} - 4 \log \frac{\sigma_{\mathrm{H} \beta}}{1000 \mathrm{~km} \mathrm{~s}^{-1}} \\
		& + (0.76 \pm 0.08) R_{\mathrm{Fe}} + 0.09
	\end{aligned}
\end{equation}

The errors in \mbh\ and \mdot\ are calculated using error propagation formulas, considering the presence of systematic errors. For the Equation \eqref{eq2}, the intrinsic scatter of $\log R_{\rm BLR}(\rm \hb)$ is about 0.17 dex \citep{yu2020extendedRL}. Systematic errors of \mbh\ and \mdot\ are adopted as 0.3 dex and 0.6 dex, respectively \citep{liu2022bian}. 

Figure \ref{fig6} shows the distributions of \mbh\ and \mdot\ calculated from Equations (\ref{eq4}), (\ref{eq8}), (\ref{eq5}), and (\ref{eq9}) for the high-$z$ sample (orange), and the low-$z$ sample of \cite{hu2008hbeta} (blue). 
For the high-$z$ sample, the mean values of $\log(\mbh/\msun)$ calculated from $\rm FWHM_{\rm \hb}$ and \shb\ are 9.32 and 9.69, with standard deviations of 0.39 and 0.37, respectively. For \mdot, the distributions are $0.54\pm 0.73$, $-0.2\pm 0.68$, respectively. The black hole mass calculated from \shb\ is 0.37 dex larger than that calculated from $\rm FWHM_{\rm \hb}$. 
The \mdot\ calculated from \shb\ is approximately 0.74 dex smaller than that calculated from $\rm FWHM_{\rm \hb}$.
For the high-$z$ quasar sample, this 0.37 dex offset is slightly larger than that for the low-$z$ SDSS subsample (by 0.25 dex) and that for the low-$z$ RM subsample (by 0.23 dex), but less than that for a sample of narrow-line Seyfert 1 galaxies (by 0.5 dex), as suggested by \cite{liu2022bian} \citep[also see][]{Bian2008}.

About the accretion strength, apart from the dimensionless accretion rate \mdot, the Eddington ratio \leddR\ is also used, where $\mdot\ \equiv \dot{M}/\dot{M}_{\rm Edd}$, $\dot{M}_{\rm Edd}=L_{\rm Edd}/c^2$.  
We use a correction factor to calculate \lb\ from \lv, $L_{\rm Bol}=BC_{\rm 5100} {L}_{\rm 5100}$, where the factor $ BC_{5100} = 53 - \log (L_{5100} / \ergs)$ \citep{marconi2004local,netzer2013agnbook,Ma2026}. 
There is a strong relation between \leddR\ and $\rm FWHM_{\hb}$-based \mdot. The best linear fit yields $\log(\leddR)=(0.50\pm 0.004)\log(\mdot)-(0.41\pm 0.004)$. Because $\mdot \propto L_{\rm 5100}^{1.5} \mbh^{-2}$, and $\leddR \propto \lb \mbh^{-1} \propto L_{ 5100}\mbh^{-1}$, $\leddR \propto \mdot^{0.5} \lv^{0.25}$. The slope of $0.50$ is the same as out best fitting slope. It is similar for the relation between \leddR\ and $\shb$-based \mdot.

For the high-$z$ quasar sample, \cite{matthews2023gemini} reported \hb\ emission line and \rfe\ measurements, with $\rm FWHM_{\rm \hb}$ distribution of $4608 \pm 1454$ \kms\ and \rfe\ distribution of $0.54 \pm 0.40$. \mbh\ calculated from $\rm FWHM_{\rm \hb}$  and \rfe\ measured by \cite{matthews2023gemini}, using our Equation (\ref{eq4}), has a mean difference of 0.08 dex with a standard deviation of 0.2 dex. Compared to \mbh\ measured from the low-$z$ SDSS sample \citep{hu2008systematic}, the high-$z$ sample has larger \lv\ and $\rm FWHM_{\rm \hb}$, resulting in larger \mbh\ (see Figure \ref{fig4}).

It was found that there exists a relation between \mdot\ and \rfe\ in low-$z$ quasars \citep[e.g.,][]{liu2022bian}. Left panel in Figure \ref{fig7} displays the relation between \mdot\ calculated from $\rm FWHM_{\rm \hb}$ and \rfe. $r_{\rm s} = 0.741$ and $p_{\rm null} = 1.058 \times 10^{-46}$ indicate the strong positive correlation between them. The orange solid line represents the best linear fit, given by $\rfe\ = (0.94 \pm 0.06) \log \mdot+ (0.09 \pm 0.05)$. Considering systematic error of 0.6 dex in \mdot, for data binned with $\mdot=0.6$ dex, the relations fitted with a second-order polynomial, yields $\rfe\ = (0.25 \pm 0.14) + (0.43 \pm 0.08) \log \mdot\ + (0.18 \pm 0.09) (\log \mdot)^2$. 
Right panel in Figure \ref{fig7} shows the case for \mdot\ calculated from \shb. $r_{\rm s} = 0.757$ and $p_{\rm null} = 1.27 \times 10^{-49}$ indicate a strong positive correlation. The linear fit is given by $\rfe\ = (0.96 \pm 0.07) \log \mdot\ + (0.8 \pm 0.039)$. The second-order polynomial fit for 0.6 dex binned \mdot\ is $\rfe= (0.68 \pm 0.18) + (0.63 \pm 0.10) \log \mdot\ + (0.11 \pm 0.06) (\log \mdot)^2$. In right panel in Figure \ref{fig7}, we also show the result for low-$z$ sample \citep{liu2022bian}, the high-$z$ quasar sample has a steeper slope. Given the strong correlation between the two quantities, \rfe\ can be used to estimate \mdot, but with significant uncertainties.

The anti-correlation between \rfe\ and the \oiii\ \(\lambda\)5008 EW, as well as the correlation between \rfe\ and \mdot\ can be understood in the context of the SED-driven wind model \citep{richards2011unification}. A softer SED (higher \mdot) enhances the wind, thereby increasing the wind density. The denser wind creates more extensive partially ionized zones where \feii\ emission originates, while the weaker X-ray radiation suppresses over-ionization of iron. Consequently, more $\rm Fe^{+}$ ions are retained, strengthening \feii\ emission and raising \rfe. Simultaneously, the enhanced wind filters the ionizing continuum, reducing the photon flux that reaches the narrow-line region and thus weakening \oiii.

\subsection{\civ\ BAL Fraction in the High-$z$ Quasar Sample}

\begin{figure}
	\centering
			\includegraphics[angle=0,width=3.5in]{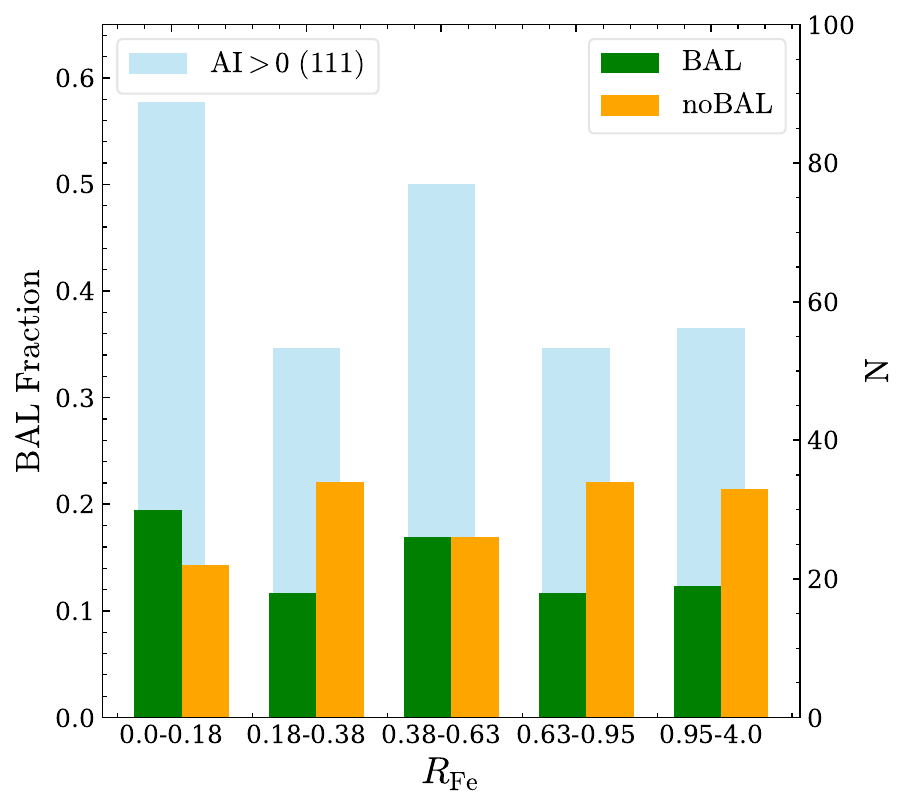}
			\caption{BAL quasar fraction ($\rm AI > 0$) across five equal-number \rfe\ bins. The left y-axis shows the BAL fraction (light blue bars), while the right y-axis shows the counts of BAL (green) and non-BAL (orange) quasars.}
	\label{fig8}
\end{figure}

Quasars often exhibit very strong and distinct broad absorption lines in the rest-frame UV spectra \citep[e.g.,][]{gibson2009catalog}. We adopt the traditional Balnicity Index (BI), adjusted absorption  index (AI), and the probability parameter $\rm  BAL_{\rm prob}$ to  indicate the reliability of \civ\ BAL quasars \citep{lyke2020sloan}. The different velocity integration ranges used in BI and AI calculations affect the identification of BAL quasars. BI integrates from 3000 \kms\ to the blue of the line center, potentially ignoring low-speed broad absorption troughs near the emission line center. In contrast, AI integrates from 0 \kms, and using $\rm AI>0$ can largely compensate for the low-velocity trough missed by BI. 

$\rm  BAL_{\rm prob}$ combines BI and AI and takes several discrete values. When BI exceeds ten times its uncertainty, $\rm  BAL_{\rm prob}=1$; when the ratio of BI to its uncertainty is less than 10 but the ratio of AI to its uncertainty exceeds 10, $\rm  BAL_{\rm prob}=0.95$; when BI equals 0 but the ratio of AI to its uncertainty remains greater than 10, $\rm  BAL_{\rm prob}=0.9$. Other quasars with a less significant BI measurement were assigned $\rm  BAL_{\rm prob}=0.75$, and quasars with a BI = 0 and a less significant AI value were assigned $\rm  BAL_{\rm prob}=0.5$. The remaining quasars were assigned $\rm  BAL_{\rm prob}=0$ \citep{lyke2020sloan}. Combining $\rm  BAL_{\rm prob}$ with BI and AI provides a more robust identification of BAL quasars. In the high-$z$ quasar sample, 111 sources have $\rm AI>0$, accounting for 43\% of the total sample. For each source with $\rm BI>0$ also have $\rm AI > 0$. Among the 111 sources, one has $ \rm BAL_{\rm prob}=0.5$, while the remaining 110 sources have $\rm  BAL_{\rm prob} \geq 0.9$.

Figure \ref{fig8} illustrates the fraction of $\rm AI>0$ quasars in equal-number \rfe\ bins. The light blue bars show the BAL fraction. The right y-axis shows the counts of BAL (green) and non-BAL (orange) quasars in each bin. The figure shows that the fraction of AI-identified BAL quasars is the highest, exceeding 50\%, at low \rfe\ values ($0$--$0.18$) and mid-range values ($0.38$--$0.63$). The fraction of \civ\  BAL quasars in the high-$z$  sample is higher than that reported by \cite{gibson2009catalog} for the SDSS DR5 quasar sample, where it is less than 15\%. 
This higher BAL fraction suggests that near-infrared selection preferentially identifies BAL quasars, particularly  low-velocity systems ($\rm AI>0$).

\subsection{Relationship between \civ\ Emission-Line Outflows and Black Hole Accretion Process}

\begin{figure*}
	\centering
	\includegraphics[width=3.2in]{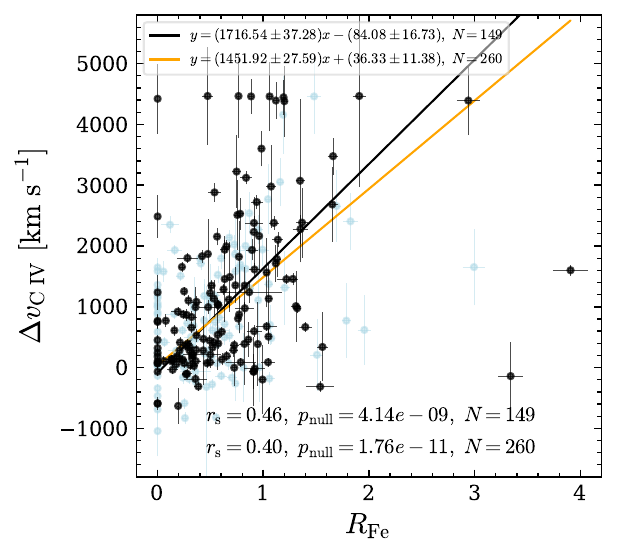}
	\includegraphics[width=3.0in]{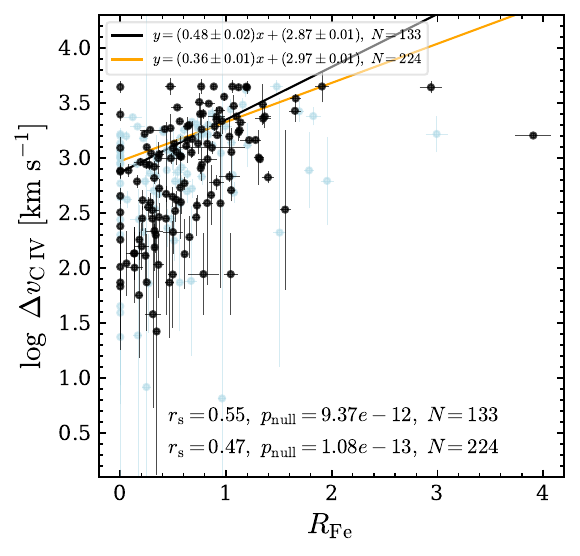}
	\caption{Left: \civ\ velocity shift ($\Delta v_{\rm \civ}$) vs. \rfe\ for the full sample of 260 quasars. The sample comprises 111 BAL quasars (light blue dots) and 149 non-BAL quasars (black dots). Right: \civ\ outflow velocity vs. \rfe\ for the 224 quasars with $\Delta v_{\rm \civ} >0 $, comprising 91 BAL quasars (light blue dots) and 133 non-BAL quasars (black dots). The orange line shows the best linear fit for all quasars and the black line shows the best linear fit for quasars excluding BAL quasars. $r_{\rm s}$ and $p_{\rm null}$ are shown in the upper right corner of each panel.}
	\label{fig9}
\end{figure*}

\begin{figure*}
	\centering
	  \includegraphics[width=2.2in]{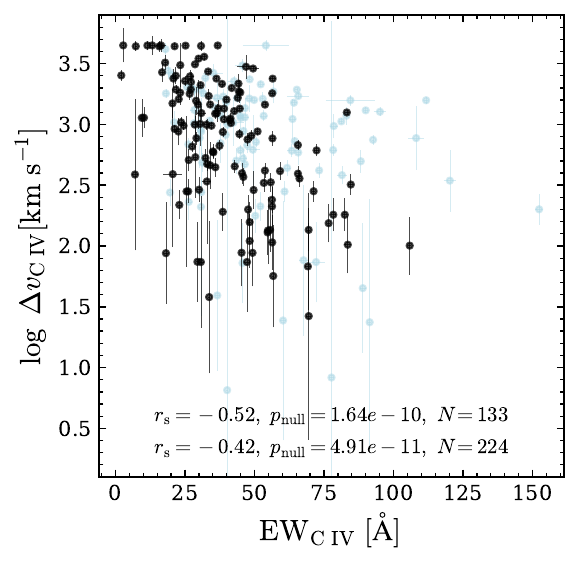}
	\includegraphics[width=2.2in]{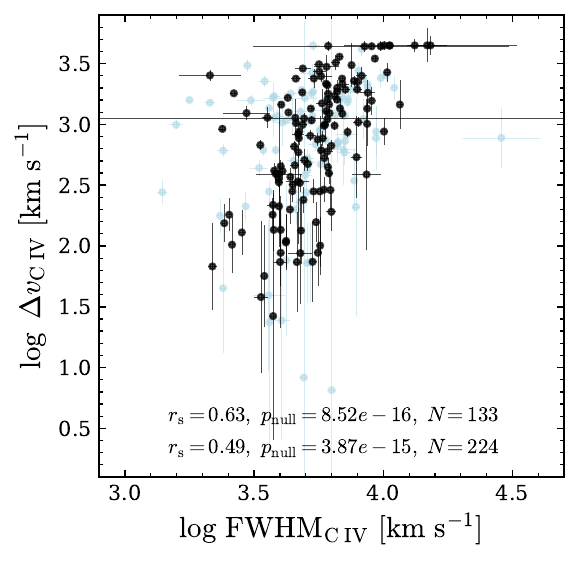}
	\includegraphics[width=2.24in]{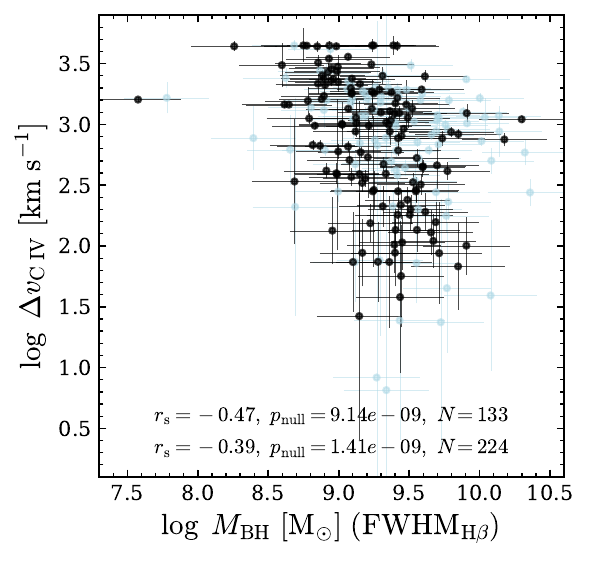}
	\caption{\civ\ outflow velocity vs. \civ\ EW (left), \civ\ FWHM (middle), and \hb-based \mbh\ for 224 quasars with $\Delta v_{\rm \civ} >0 $. The sample comprises 91 BAL quasars (light blue dots) and 133 non-BAL quasars (black dots). $r_{\rm s}$ and $p_{\rm null}$ are reported at the bottom of each panel.}
	\label{fig10}
\end{figure*}

For the SDSS spectra of 260 high-$z$ quasars, we mask BAL data points during \civ\ emission line fitting to obtain reliable measurements of the \civ\ emission line peak wavelength.
The velocity shift of the \civ\ line relative to the systematic rest frame is given by:
\begin{equation}\label{eq:vshift}
	\Delta v_{\civ} = (1549\,\text{\AA} - \lambda_{\rm peak}) \times \frac{c}{1549\,\text{\AA}}
\end{equation}
where $\lambda_{\rm peak}$ is the fitted \civ\ rest-frame wavelength and $c$ is the speed of light.
The uncertainty in $\Delta v_{\civ}$ includes contributions from both the peak wavelength measurement and the 
systemic redshift error, combined in quadrature. The systematic redshift errors for \oiii, \mgii, and \hb\ line are 50 \kms, 200 \kms, and 400 \kms, respectively \citep{shen2016sloan}. Of the 260 quasars, 224 ($\sim 87\%$) have $\Delta v_{\civ}>0$ (i.e., blueshift \civ), while 36 ($\sim 13\%$) are redshifted (left panel in Figure \ref{fig9}). 

Figure \ref{fig9} shows the distribution of UV \civ\ emission-line velocity shifts versus optical \rfe. The velocity shift ranges from about $-1000$ to $4500$ \kms. Positive $\Delta v_{\civ}$ indicates blueshifted \civ\ emission due to outflows, while negative $\Delta v_{\civ}$ indicates redshift due to inflow. The left panel presents the full sample in Figure~\ref{fig9}, where the Spearman correlation coefficient $r_{\rm s} = 0.40$ with $p_{\rm null} = 1.76 \times 10^{-11}$ indicates a moderate positive correlation. For the 224 \civ\ outflow quasars with $\Delta v_{\civ}>0$, the right panel reveals a stronger positive correlation, with $r_{\rm s} = 0.47$  and $p_{\rm null} = 1.08 \times 10^{-13}$. A linear fit to the outflow subsample, accounting for the errors in both variables, yields $\log (\Delta v_{\civ}/\kms) = (0.36 \pm 0.01) \rfe\ + (2.97 \pm 0.01)$. Excluding BAL quasars strengthens these correlations further: for the full sample,$r_{\rm s}$ increases from 0.40 to 0.46, and for the outflow subsample, from 0.47 to 0.55.

The left panel of Figure \ref{fig10} shows \civ\ outflow velocity versus \civ\ EW for 224 quasars with $\Delta v_{\civ}>0$. A moderate negative correlation is found ($r_{\rm s} = -0.42$, $p_{\rm null} = 4.91 \times 10^{-11}$). Excluding 91 BAL quasars ($\rm AI > 0$), $r_{\rm s} = -0.52$ and $p_{\rm null} = 1.64 \times 10^{-10}$ show a stronger negative correlation (see left panel of Figure \ref{fig10}). This is consistent with the results of \citet{richards2011unification} and \citet{ge2019blueshift}, who found that quasars with larger \civ\ EWs tend to show smaller outflows velocities.
The middle panel of Figure \ref{fig10} shows \civ\ outflow velocity versus $\rm FWHM_{\civ}$. A strong positive correlation is found ($r_{\rm s} = 0.49$,  $p_{\rm null} = 3.87 \times 10^{-15}$).
After removing BAL quasars, the correlation strengthens further ($r_{\rm s} = 0.63$, $p_{\rm null} = 8.52 \times 10^{-16}$; see the middle panel of Figure \ref{fig10}). Quasars with larger $\rm FWHM_{\civ}$ tend to show larger \civ\ emission-line outflows velocities.
For the relations between the \civ\ outflow velocity and the continuum luminosities \(L_{\rm 1700}\) and \(L_{\rm 5100}\), as well as \(\rm FWHM_{\hb}\), the Spearman correlation test yields no significant correlations with $L_{1700}$ ($r_{\rm s} = 0.12$, $p_{\rm null} = 0.08$) or $L_{5100}$ ($r_{\rm s} = 0.10$, 
$p_{\rm null} = 0.14$). However, a weak negative correlation is 
found with $\mathrm{FWHM}_{\mathrm{H}\beta}$ ($r_{\rm s} = -0.24$, 
$p_{\rm  null} = 2.30 \times 10^{-4}$).
It is not consistent with the weak relation between the \civ\ blueshift and \lv\ found by \cite{ge2019blueshift}.
For the relation between \(\rm FWHM_{\civ}\) and \(\rm FWHM_{\hb}\), $r_{\rm s} = -0.08$ and \(p_{\rm null} = 0.24\), indicating no significant correlation. Excluding BAL quasars, the remaining 133 non-BAL sources also show no strong correlations.
The right panel of Figure \ref{fig10} shows the relationship between \civ\ outflow velocity and \mbh. A moderate negative correlation is found ($r_{\rm s} = -0.39$, $p_{\rm null} = 1.41 \times 10^{-9}$). For the 133 non-BAL quasars, the correlation becomes stronger ($r_{\rm s} = -0.47$, $p_{\rm null} = 9.14 \times 10^{-9}$). For \shb-based \mbh, the correlation becomes weaker: $r_{\rm s} = -0.24$, $p_{\rm null} = 2.18 \times 10^{-4}$ for the 224 quasars with outflows, and $r_{\rm s} = -0.24$, $p_{\rm null} = 6.24 \times 10^{-3}$ for the 133 non-BAL quasars.

Given the strong correlation between \mdot\ and \leddR, we adopt \mdot\ as a proxy for accretion strength to investigate the radiative-driving origin of \civ\ outflows. Figure \ref{fig11} shows the relation between the outflow velocity and $\rm FWHM_{\hb}$-based \mdot. The Spearman correlation test yields a moderate positive correlation ($r_{\rm s} = 0.44$, $p_{\rm null} = 8.31 \times 10^{-12}$), consistent with \cite{ge2019blueshift}. A linear fit gives $\log (\Delta v_{\rm \civ} /\kms)= (1.49 \pm 0.23) \times  \log(\mdot) + (2.06 \pm 0.15)$. For \shb-based \mdot, the correlation weakens ($r_{\rm s} = 0.30$, $p_{\rm null} = 5.06 \times 10^{-6}$).

The correlation analysis between \civ\ emission-line properties and other measured quantities (particularly \civ\ and \hb\ FWHMs and continuum luminosities at 1700 \AA\ and 5100 \AA) reveals the complex and irregular nature of the \civ\ profiles in this sample. When calibrating the \mbh\ estimator based on \(\rm FWHM_{\civ}\) against the \hb-based \mbh, it is necessary to account for \civ\ emission-line outflows to reduce the dispersion, as well as the \civ\ EW \citep[e.g.,][]{shen2012, bian2012, coatman2017, Dix2023}.

The moderate positive correlation between the \civ\ emission-line blueshift and \mdot\ found in our high-$z$ sample can be naturally understood in the framework of the radiation-pressure-driven disk-wind model \citep[e.g.,][]{Murray1995,Proga2000,richards2011unification}. In this scenario, a higher \mdot\ corresponds to a softer SED (more negative $\alpha_{\rm ox}$), where the relative enhancement of UV photons increases the radiation line-driving force, while the weaker X-ray flux reduces over-ionization and preserves more effective \civ\ scatterers \citep{Leighly2004, Casebeer2006}. Consequently, the wind is accelerated to higher terminal velocities, producing larger \civ\ blueshifts. 
BAL quasars are consistent with being dominated by the wind component in a two-component disk-wind model \citep{richards2011unification}. The high BAL fraction (43\%) in our NIR-selected sample further supports this interpretation, suggesting that our selection preferentially targets quasars with softer SEDs and stronger winds, as expected from the  \mdot-dependent disk-wind model. The results presented in this work are summarized in Table~\ref{table2}.

\subsection{Radiation Pressure Driven \civ\ Emission-Line Outflow}

\begin{figure*}
	\centering
	\includegraphics[width=3.2in]{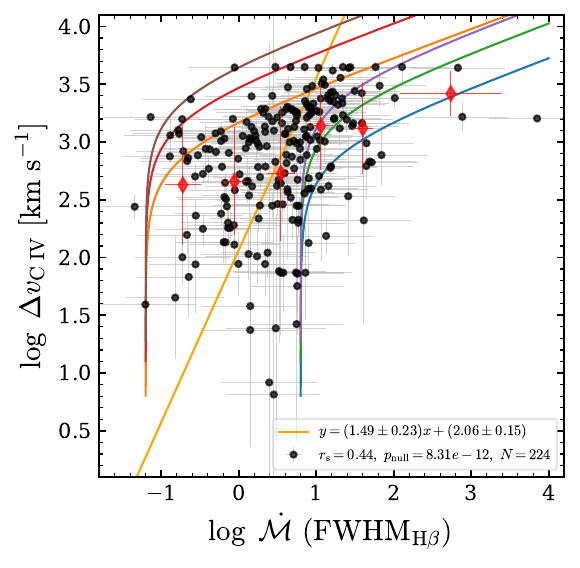}
	\includegraphics[width=3.2in]{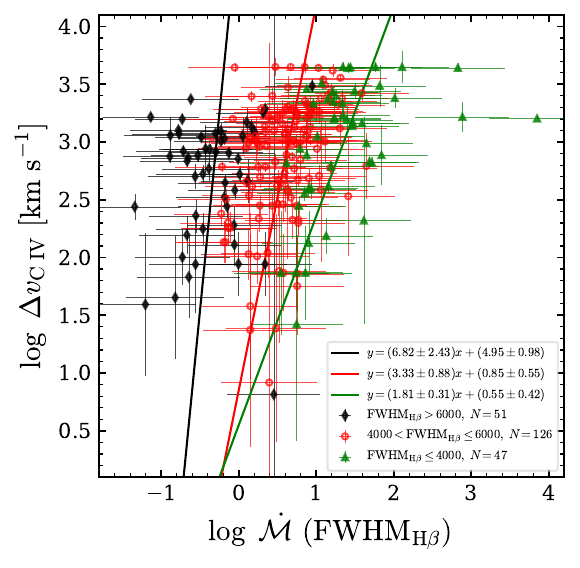}
	\caption{Left: \civ\ outflow velocity vs. $\rm FWHM_{\hb}$-based \mdot\  for 224 high-$z$ quasars. The orange line shows the best linear fit. The red rhombuses indicate binned data in 0.6 dex intervals of \mdot.
$r_{\rm s}$ and $p_{\rm null}$ are shown in the bottom right corner. 
The blue, green, and purple curves show the theoretical predictions for $M = 1$ with $\rm FWHM_{\hb}=$ 2000, 4000, and 6000 \kms, respectively; the orange, ochre, and brown curves show the theoretical predictions for $M = 10$ with the same $\rm FWHM_{\hb}$ values.
Right: \civ\ outflow velocity vs. $\rm FWHM_{\hb}$-based \mdot\ for three $\rm FWHM_{\hb}$ bins: $\leq 4000$ \kms\ (green), $4000$--$6000$ \kms\ (red), and $> 6000$ \kms\ (black). The best linear fits are shown in the bottom right corner.}
	\label{fig11}
\end{figure*}

\begin{figure*}
	\centering

	\includegraphics[width=3.2in]{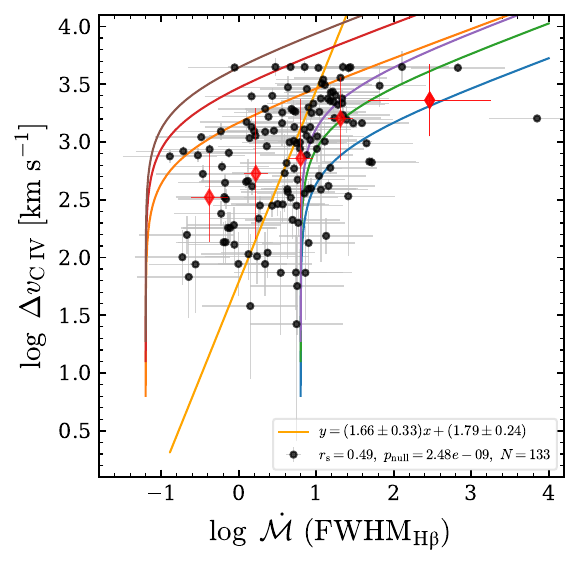}
	\includegraphics[width=3.2in]{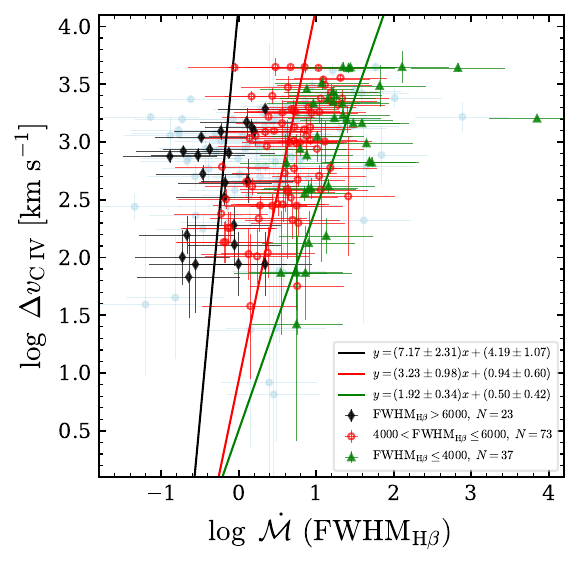}
	\caption{Left: The relationship between the outflow velocity magnitude and \(\dot{\mathcal{M}}\) for 133 non-BAL quasar. Right:  \civ\ outflow velocity vs. $\rm FWHM_{\hb}$-based \mdot\ for three $\rm FWHM_{\hb}$ bins. BAL quasar data points are shown as light blue background points for comparison. Axes and curves are as in Figure \ref{fig11}; fit parameters are listed in the legend.}
	\label{fig12}
\end{figure*}

Quasar gas immersed in an intense radiation field can give rise to large-scale outflows under specific conditions; these outflows are referred to as winds. The acceleration of a cloud of mass $M_{\mathrm{c}}$ with density $\rho$ can be  derived from the general equation of motion, as described by \citet{netzer2013agnbook} and \citet{liu2022bian},
\begin{equation}\label{eq:motion}
	a(r) = a_{\mathrm{rad}}(r) - g(r) - \frac{1}{\rho} \frac{d P}{d r} + \frac{f_d}{M_c}
\end{equation}
where $g(r)$ is the gravitational acceleration, $a_{\rm rad}(r)$ is the radiative pressure acceleration, and $f_{\rm d}$ is the drag force. Neglecting the gas pressure gradient and the drag force for pure wind flows, 
\begin{equation}\label{eq:motion_a}
\begin{aligned}
	a(r)=a_{ rad}(r) - g(r) & \approx \frac{\sigma_T L}{4 \pi r^2 \mu m_H c} \left[M(r) - \frac{L_{\mathrm{Edd}}}{L}\right]  \\
	&= \frac{G}{r^2} M_{\mathrm{BH}} \left[M(r) \frac{L}{L_{\mathrm{Edd}}} - 1\right]
	\end{aligned}
\end{equation}
where \(L\) is the total luminosity, which can be approximated as \lb; $M(r)$ is the multiplier for radiative pressure, the ratio of the total radiative pressure to the pressure from Compton scattering, with $M(r) \geq 1$, and for fully ionized gas $M(r) = 1$. Assuming radiative acceleration of the \civ\ emission-line outflow from rest at $r_1$ to velocity $v$ at $r_2$ within the \civ\ broad-line region, the outflow velocity is:
\begin{equation}
	v^2 = 2 G M_{\mathrm{BH}} \left(1 - M \frac{L}{L_{\mathrm{Edd}}}\right) \left(\frac{1}{r_2} - \frac{1}{r_1}\right)
\end{equation}
where we assume $r_2 =n r_1$ with $n > 1$, and  $M(r) \frac{L}{\ledd}-1$ is considered a constant.
Using the single-epoch virial black hole mass $\mbh= f {\rm FWHM}_{\rm \hb}^2 r_1 /G $, we obtain:
\begin{equation}
	v^2 \approx 2 \frac{n-1}{n} f \mathrm{FWHM}_{\rm H \beta}^2 \left(M(r) \frac{L}{L_{\mathrm{Edd}}} - 1\right)
\end{equation}
Since the force multiplier $M(r)$ is related to the ionization parameter \citep{chelouche2005dynamical}, assuming $M(r) = M_0 \left(L / L_{\mathrm{Edd}}\right)^{-\alpha}$ in the above equation leads to the outflow velocity:
\begin{equation}\label{eq14}
	v^{2} \approx 2 \frac{n-1}{n} f \mathrm{FWHM}_{\rm H \beta}^2 \left[M_0 \left(L / L_{\mathrm{Edd}}\right)^{1-\alpha} - 1\right]
\end{equation}

Assuming $n = 1.1$, $f=1$, and a force multiplier $M(r)$ independent of the Eddington ratio, with $\log(\leddR)= 0.5 \log \mdot\ - 0.4$, theoretical curves are drawn for six cases: $\rm FWHM_{\hb}=$ 2000, 4000, and 6000 \kms, each with $M=1$ and $M=10$. In Figure \ref{fig11}, the blue, green, and purple curves correspond to $M = 1$ with $\rm FWHM_{\hb}=$ 2000, 4000, and 6000 \kms, respectively; the orange, ochre, and brown curves correspond to $M = 10$ with the same $\rm FWHM_{\hb}$ values.

The six theoretical curves in Figure \ref{fig11} show that the absence of very high-speed outflows suggests a short acceleration region scale ($n$ slightly greater than 1), while high-speed outflows require either a larger force multiplier $M(r)$ or a larger acceleration scale (larger $n$). The theoretical predictions are consistent with most of the observed outflows in the high-$z$ quasar sample.

From Equation \eqref{eq14}, $\rm FWHM_{\hb}$ is associated with the \civ\ outflow velocity. Based on $\rm FWHM_{\hb}$, we divide the 224 high-$z$ quasars into three bins: $\rm FWHM_{\hb} \leq 4000$ \kms, $4000 < \rm FWHM_{\hb} \leq 6000$ \kms, and $\rm FWHM_{\hb} \geq 6000$ \kms. 
For quasars with $\rm FWHM_{\hb} > 6000$ \kms, $r_{\rm s} = 0.11$ and $p_{\rm null} = 0.43$. 
The best linear fit is  $\log \Delta v_{\civ} = (6.82 \pm 2.43) \log \mdot\ + (4.95 \pm 0.98)$.
For quasars with $4000 < \rm FWHM_{\hb} \leq 6000$ \kms, $r_{\rm s} = 0.44$ and $p_{\rm null} = 2.82 \times 10^{-7}$. 
The best linear fit is  $\log \Delta v_{\civ} = (3.33 \pm 0.88) \log\mdot\ + (0.85 \pm 0.55)$.
For quasars with $\rm FWHM_{\hb} \leq 4000$ \kms, $r_{\rm s} = 0.54$ and $p_{\rm null} = 7.97 \times 10^{-5}$.
The best linear fit is  $\log \Delta v_{\civ}  = (1.81\pm 0.31) \log\mdot\ + (0.55 \pm 0.42)$.
The six theoretical curves indicate that for the same \mdot, a larger $\rm FWHM_{\hb}$ results in a higher \civ\ outflow velocity, and that a larger \mdot\ and $\rm FWHM_{\hb}$ both lead to a high-speed \civ\ outflows. Excluding BAL quasars, the results for the 133 non-BAL quasars are consistent with the full outflow sample (see Figure \ref{fig12}), which implies that BALs and non-BALs have the same underlying origin of \civ\ emission traced outflow \citep{Rankine2020}.

\section{Conclusions}\label{s4}

For a sample of 260 high-$z$ quasars ($1.5 < z < 3.6$) with optical spectra from SDSS (3800 -- 9200 \AA) and near-infrared spectra from GNIRS (8000 -- 25000 \AA) in the observed-frame, we investigate the relation between \civ\ emission-line outflows and SMBH accretion. The main conclusions are as follows:

\begin{itemize}

\item 
We uniformly perform spectral fitting of the optical and UV spectra in the rest frame for 260 high-$z$ quasars ($1.5<z<3.6$) to measure emission-line properties, including the optical parameters $\rfe, ~\rm{FWHM}_{\hb}$, and \shb, as well as the UV \civ\ peak wavelength, and compare the optical properties with low-$z$ samples from \cite{hu2008systematic} ($z<0.8$). The measured $\rm FWHM_{\hb}$ and \shb\ are $5252 \pm1525$ \kms\ and $3395 \pm 774$ \kms, respectively, larger than those in the low-$z$ SDSS sample. The optical \rfe\ and $D_{\rm \hb}$ are $0.61 \pm 0.55$ and $1.58 \pm 0.41$, respectively, smaller than the values for the low-$z$ samples from \cite{hu2008systematic}. Larger \shb\ and smaller $D_{\rm \hb}$ imply stronger wing components in the \hb\ emission-line profiles for the high-$z$ quasar sample. 
The $\log \mbh$ and $\log \mdot\ $ from $\rm FWHM_{\rm \hb}$ are $9.32  \pm 0.39$ and $0.54\pm 0.73$, respectively, while those from \shb\ are $9.69 \pm 0.38$ and $-0.20 \pm 0.68$, respectively.

\item  
It is found that optical \rfe\ is moderately anti-correlated with \oiii\ EW ($r_{\rm s} = -0.52$, $p_{\rm null} = 1.045 \times 10^{-19}$) and strongly correlated with the $\rm FWHM_{\hb}$-based \mdot\ ($r_{\rm s} = 0.74$, $p_{\rm null} = 1.06 \times 10^{-46}$), as in the low-$z$ sample.
For high-$z$ quasars, the optical \rfe\ can be used as a tracer of \mdot, as is the case for low-$z$ quasars. Accounting for systematic errors in \mdot, we bin the sample in 0.6 dex intervals of \mdot\ and perform a second-order polynomial fit: $\rfe\ = (0.25 \pm 0.14) + (0.43 \pm 0.08) \log \mdot+ (0.18 \pm 0.09) (\log \mdot)^2$. 
The anti-correlation between \rfe\ and the \oiii\ $\lambda5008$ EW, together with the correlation between \rfe\ and \mdot, can be understood in the context of the SED-driven wind model \citep{richards2011unification}. A higher \mdot\ corresponds to a softer SED, which enhances the radiation-pressure-driven wind. This increases the wind density and strengthens the \feii\ emission (higher \rfe), while simultaneously filtering the ionizing continuum. The reduced ionizing photon flux reaching the narrow-line region weakens the \oiii\ emission, giving rise to the observed anti-correlation.

\item  
In the high-$z$ quasar sample with $H < 16.5$, 111 quasars ($\sim 43\%$) have $\rm AI > 0$. Compared to SDSS DR5 quasar sample of \cite{gibson2009catalog}, the BAL quasars fraction is larger, indicating that the near-infrared selection preferentially identifies BAL quasars, mostly low-speed system ($\rm AI>0$).  
Of the 260 high-$z$ quasars, 224 ($\sim 87\%$) have $\Delta v_{\civ}>0$ (i.e., blueshift \civ), while 36 ($\sim 13\%$) are redshifted. 
Moderate positive correlations are found between UV \civ\ blueshift (relative to systemic redshift) and both the optical \rfe\ and \mdot, regardless of whether BAL quasars are excluded. BALs and non-BALs have the same underlying origin of \civ\ emission traced outflow. 
This positive correlation can be naturally understood in the framework of the radiation-pressure-driven disk-wind model. A higher \mdot\ corresponds to a softer SED, where the relative enhancement of UV photons increases the radiation line-driving force, while the weaker X-ray flux reduces over-ionization and preserves more effective \civ\ scatterers. Consequently, the wind is accelerated to higher terminal velocities, producing larger \civ\ blueshifts. 
According to the radiation pressure-driven disk-wind model, \civ\ outflows can be explained and are related to $\rm{FWHM}_{\hb}$ and \mdot.

\end{itemize}

\section*{Data Availability}
The sample used in this study is available at \url{https://datalab.noirlab.edu/gnirs_dqs.php}. The data used in this work and the spectral fitting figures are available at Zenodo (link to be provided upon acceptance). The data format is described in Table~\ref{table1}.

\begin{acknowledgments}
We thank the anonymous referee for the valuable comments and suggestions, which have helped us improve the manuscript. This work has been supported by the National Science Foundation of China (No. 12573008).
\end{acknowledgments}

\begin{deluxetable*}{cccc}
\tablecaption{FITS Catalog Format \label{table1}}
\tablehead{
\colhead{Column} & \colhead{Format} & \colhead{Units} & \colhead{Description}
}
\startdata
SDSS\_NAME & STRING & \nodata & SDSS J2000 designation \\
Z & DOUBLE & \nodata & Redshift \\
Z\_SYS\_ERR & DOUBLE & \nodata & Systematic redshift error \\
GNIRS\_RANGE & DOUBLE[2] & \AA & Rest-frame fitting range of GNIRS spectra \\
SDSS\_MASK & STRING & \AA & Rest-frame mask window of \ion{C}{4} absorption line \\
LOGL5100 & DOUBLE & erg s$^{-1}$ & Continuum luminosity at rest-frame 5100 \AA \\
LOGL5100\_ERR & DOUBLE & erg s$^{-1}$ & Error of LOGL5100 \\
BR\_HB\_FWHM & DOUBLE & km s$^{-1}$ & FWHM of broad H$\beta$ emission line \\
BR\_HB\_FWHM\_ERR & DOUBLE & km s$^{-1}$ & Error of BR\_HB\_FWHM \\
BR\_HB\_SIGMA & DOUBLE & km s$^{-1}$ & Line dispersion of broad H$\beta$ emission line \\
BR\_HB\_SIGMA\_ERR & DOUBLE & km s$^{-1}$ & Error of BR\_HB\_SIGMA \\
BR\_HB\_EW & DOUBLE & \AA & Rest-frame equivalent width of broad H$\beta$ emission line \\
BR\_HB\_EW\_ERR & DOUBLE & \AA & Error of BR\_HB\_EW \\
OIII\_FWHM & DOUBLE & \kms & FWHM of [\ion{O}{3}] 5008 emission line \\
OIII\_EW & DOUBLE & \AA & Rest-frame equivalent width of [\ion{O}{3}] 5008 emission line \\
RFE & DOUBLE & \nodata & Rest-frame 4434--4484 \AA\ optical Fe II flux relative to broad H$\beta$ \\
RFE\_ERR & DOUBLE & \nodata & Error of RFE \\
LOGL1700 & DOUBLE & erg s$^{-1}$ & Continuum luminosity at rest-frame 1700 \AA \\
CIV\_PEAK & DOUBLE & \AA & Rest-frame peak wavelength of \ion{C}{4} emission line \\
CIV\_PEAK\_ERR & DOUBLE & \AA & Error of CIV\_PEAK \\
CIV\_FWHM & DOUBLE & km s$^{-1}$ & FWHM of \ion{C}{4} emission line \\
CIV\_FWHM\_ERR & DOUBLE & km s$^{-1}$ & Error of CIV\_FWHM \\
CIV\_EW & DOUBLE & \AA & Rest-frame equivalent width of \ion{C}{4} \\
CIV\_EW\_ERR & DOUBLE & \AA & Error of CIV\_EW \\
BI & DOUBLE & \nodata & \ion{C}{4} BALnicity index \\
AI & DOUBLE & \nodata & \ion{C}{4} absorption index \\
BAL\_PROB & DOUBLE & \nodata & \ion{C}{4} broad absorption line probability \\
LOGMBH\_FWHM & DOUBLE & $\rm M_\odot$ & Black hole mass calculated from BR\_HB\_FWHM \\
LOGMBH\_ERR\_FWHM & DOUBLE & $\rm M_\odot$ & Error of LOGMBH\_FWHM \\
LOGMBH\_SIGMA & DOUBLE & $\rm M_\odot$ & Black hole mass calculated from BR\_HB\_SIGMA \\
LOGMBH\_ERR\_SIGMA & DOUBLE & $\rm M_\odot$ & Error of LOGMBH\_SIGMA \\
LOGMDOT\_FWHM & DOUBLE & \nodata & Dimensionless accretion rate calculated from BR\_HB\_FWHM \\
LOGMDOT\_ERR\_FWHM & DOUBLE & \nodata & Error of LOGMDOT\_FWHM \\
LOGMDOT\_SIGMA & DOUBLE & \nodata & Dimensionless accretion rate calculated from BR\_HB\_SIGMA \\
LOGMDOT\_ERR\_SIGMA & DOUBLE & \nodata & Error of LOGMDOT\_SIGMA \\
\enddata
\tablecomments{Data and spectral fitting figures are available at Zenodo (link to be provided upon acceptance).}
\end{deluxetable*}

\begin{deluxetable*}{ccccc}
    \tablecaption{Correlations between the \civ\ outflow velocity, $\Delta v_{\civ}$, and different quantities. \label{table2}}
    \tablehead{
        \colhead{} &
        \multicolumn{2}{c}{$\Delta v_{\civ}>0$ \& non-BAL} &
        \multicolumn{2}{c}{$\Delta v_{\civ}>0$} \\
        \colhead{} &
        \multicolumn{2}{c}{$N=133$} &
        \multicolumn{2}{c}{$N=224$} \\
        \hline
        \colhead{Quantities} &
        \colhead{$r_{\rm s}$} & \colhead{$p_{\rm null}$} &
        \colhead{$r_{\rm s}$} & \colhead{$p_{\rm null}$} 
    }
    \startdata
    $\rfe$ &
    $0.55$ & $9.37\times10^{-12}$ &
    $0.47$ & $1.08\times10^{-13}$ \\
    $\rm EW_{\civ}$ &
    $-0.52$ & $1.64\times10^{-10}$ &
    $-0.42$ & $4.91\times10^{-11}$ \\
    $\rm FWHM_{\civ}$ &
    $0.63$ & $8.52\times10^{-16}$ &
    $0.49$ & $3.87\times10^{-15}$ \\
    $\mbh\ \rm (FWHM_{\hb})$ &
    $-0.47$ & $9.14\times10^{-9}$ &
    $-0.39$ & $1.41\times10^{-9}$ \\
    $\mdot\ \rm (FWHM_{\hb})$ &
    $0.49$ & $2.48\times10^{-9}$ &
    $0.44$ & $8.31\times10^{-12}$ \\
    $\mbh\ \rm (\sigma_{\hb})$ &
    $-0.24$ & $6.24\times10^{-3}$ &
    $-0.24$ & $2.18\times10^{-4}$ \\
    $\mdot\ \rm (\sigma_{\hb})$ &
    $0.27$ & $1.36\times10^{-3}$ &
    $0.30$ & $5.06\times10^{-6}$ \\
    \enddata
    \tablecomments{$\rm FWHM_{\hb}$ or $\rm \sigma_{\hb}$ in parentheses indicates that \mbh\ and \mdot\ are calculated from the broad \hb\ FWHM or line dispersion, respectively.}
\end{deluxetable*}

\newpage
\bibliographystyle{aasjournal}
\bibliography{ref}
\end{document}